\documentclass{ws-procs9x6}

\usepackage{slashed}

\newcommand{\arrow}{$- \hspace{-0.18cm}>$}

\newcommand{\met}{\slashed{E}_T}
\newcommand{\calchep}{{\tt CalcHEP }}
\newcommand{\pythia}{{\tt PYTHIA }}
\newcommand{\micromegas}{{\tt micrOmegas }}
\newcommand{\comphep}{{\tt CompHEP }}

\begin{document}

%\title{CalcHEP and PYTHIA Tutorials \\
%TASI 2011: The Dark Secrets of the Terascale}

\title{TASI 2011: CalcHEP and PYTHIA Tutorials}

\author{Kyoungchul Kong}

\address{
Department of Physics and Astronomy, \\ University of Kansas, \\
Lawrence, KS 66045 USA \\
E-mail: kckong@ku.edu}

\begin{abstract}
This note summarizes a pedagogical tutorial on \calchep and \pythia that was given at TASI 2011 program.
\end{abstract}


\bodymatter

%%%%%%%%%%%%%%%%%%%%%%%%%%%%%%%%%%%%%%%%%%%%%%%%%%
\section{Introduction}
%%%%%%%%%%%%%%%%%%%%%%%%%%%%%%%%%%%%%%%%%%%%%%%%%%

Analysis in high energy physics and astrophysics is becoming increasingly complex. 
As part of the theoretical efforts tackling this complication, many multi-purpose event generators are developed.
TASI program in 2011 provided tutorials on selected tools during the school:  
{\tt CalcHEP}\cite{Pukhov:2004ca,Pukhov:1999gg,Belyaev:2000wn,Belyaev:2012qa}, {\tt PYTHIA}\cite{Sjostrand:2007gs,Sjostrand:2006za}, 
{\tt MCFM}\cite{mcfm}, {\tt PGS}\cite{pgs} and {\tt micrOmegas} \cite{Belanger:2010gh,Belanger:2008sj}. 
There are many other useful tools available and users are strongly encouraged to experience those and compare with 
selected programs. Each tool has its advantages and disadvantages and users should be aware of their limitations. 
A Repository for Beyond-the-Standard-Model Tools can be found from several sources, some of which are listed below. 
	\begin{enumerate}
	\item http://www.ippp.dur.ac.uk/montecarlo/BSM/ 
	\item MC4BSM workshop: http://theory.fnal.gov/mc4bsm/ 
	\end{enumerate}
This note provides a quick summary for how to use \calchep and \pythia and emphasizes the use of batch modes that are often ignored. 
Advanced users should refer to manuals \cite{Belyaev:2012qa,Belyaev:2000wn,Sjostrand:2007gs,Sjostrand:2006za,Pukhov:2004ca,Pukhov:1999gg}. 
All plots and source codes are available from http://susy.phsx.ku.edu/$\sim$kckong/tasi/.
Useful exercise for \calchep and \pythia are also found from PiTP school:  http://www.phys.ufl.edu/$\sim$matchev/pitp/ as well as their own web pages.

%%%%%%%%%%%%%%%%%%%%%%%%%%%%%%%%%%%%%%%%%%%%%%%%%%
\section{CalcHEP}
%%%%%%%%%%%%%%%%%%%%%%%%%%%%%%%%%%%%%%%%%%%%%%%%%%

\calchep is a package for evaluation of Feynman diagrams, integration over multi-particle phase space, and 
event generation.
It is a menu-driven system with contextual help. 
The notations used in \calchep are very similar to those used in particle physics.
The \calchep package consists of two parts: symbolic and numerical.
Both parts are written in the C programming language. 
The symbolic part produces C codes for a squared matrix element, and they are used in the numerical calculation later on. 
We summarize some features of \calchep below.

\begin{itemize}

\item \calchep stands for Calculators for High Energy Physics. %and authors are A. Pukhov, A. Belyaev and N. Christensen.
%	\begin{itemize}
%	\item It is a package for evaluation of Feynman diagrams, integration over multi-particle phase space, and event generation.
%	\item Webpage: http://theory.sinp.msu.ru/$\sim$pukhov/calchep.html.
%	\item Authors are A. Pukhov, A. Belyaev and N. Christensen.
%	\end{itemize}

\item \comphep is a very similar program and shares many codes with \calchep 
	\begin{itemize}
	\item Download from http://comphep.sinp.msu.ru/.
	\item Current version is written in 2010.
	\end{itemize}
	
\item Features
	\begin{itemize}
	\item \calchep can evaluate any decay and scattering processes within any (user defined) model.
	\item It has an easy user interface, and keeps correct spin correlations.
	\item Symbolic calculation makes analysis very easy for $1 \to 2$ and $2 \to 2$ processes.
	\item It is easy and quick to get plots from a model.
	\item It is linked to \micromegas for dark matter study.
	\end{itemize}

\item Limitations
	\begin{itemize}
	\item \calchep deals with tree-level processes only in the squared matrix element calculation.
%	\item Spin and polarization are summer over for external particles.
	\item It provides spin/polarization averaged amplitudes.
	\item Limit on the number of external legs (involved particles) is 8 (for example, $2 \to 6$ or $1 \to 7$), and 
	there is a limit on the number of diagrams.	
	\end{itemize}

\item CalcHEP is especially useful
	\begin{itemize}
	\item when users need quick estimation of cross sections and decay widths (at tree-level),
	\item when users cross-check calculations of other tools,
	\item when users need to evaluate relic abundance of dark matter and direct/indirect detection limit,
	\item when processes are relatively short.
	\end{itemize}

\end{itemize}

%%%%%%%%%%%%%%%%%%%%%%%%%%%%%%%%%%%%%%%%%%%%%%%%%%
\subsection{Installation}
%%%%%%%%%%%%%%%%%%%%%%%%%%%%%%%%%%%%%%%%%%%%%%%%%%

\calchep can be downloaded from 
\begin{verbatim}
http://theory.sinp.msu.ru/~pukhov/calchep.html.
\end{verbatim}
To install, 
\begin{enumerate}
\item unpack: tar -zxvf ${\rm calchep\_3.0.tar.gz}$
\item install by typing ``make''.
\item If you encounter a problem with ``blind mode", you need to install libx11-dev.
\item create your own working directory: (e.g.) mkUsrDir WORK
\item to run, go to the WORK directory and simply execute ``./calchep" 
\end{enumerate}
The following subdirectories are created under the {\tt WORK} directory: {\tt bin, models, results, tmp}.
Each model (under ``models'' directory) is defined in terms of 5 files. We will look into those in the next section.
Other than main source codes, \$CALCHEP/bin and  \$CALCHEP/utile contain useful scripts and codes. 
For more information see ``README'' in each directory. 
We will go over some useful routines with examples.

\calchep is very easy to use. 
All selections can be made with ``ENTER'' or ``RETURN'' key.
To go to the previous menu, simply press ``ESC'' key. 
If there are any questions, users can get real-time help by pressing ``F1''. 

%%%%%%%%%%%%%%%%%%%%%%%%%%%%%%%%%%%%%%%%%%%%%%%%%%
\subsection{A Closer look into model files: \\
Particles, Vertices, Parameters, Constraints, Libraries}
%%%%%%%%%%%%%%%%%%%%%%%%%%%%%%%%%%%%%%%%%%%%%%%%%%

A physics model is defined by a set of files with each containing a different aspect of the model: 
``prtcls\#.mdl'' for particle definition, 
``vars\#.mdl'' for independent parameters,
``func\#.mdl'' for dependent parameters, 
``lgrng\#.mdl'' for interactions, and 
``'extlib\#.mdl" for external routines. 
\begin{enumerate}
\item ``{\tt Particles}": particles are defined in this file. For example, particle definition for the Standard Model is shown below 
(usually ``prtcls1.mdl'').
{\scriptsize 
\begin{verbatim}
Standard Model
 Particles 
Full  name |>A <|>A <|number|2*spin|mass|width|color|aux|>LaTex(A)<|>LaTeX(A+)   <|
gluon      |G   |G   |21    |2     |0   |0    |8    |G  |g         |g
photon     |A   |A   |22    |2     |0   |0    |1    |G  |\gamma    |\gamma
Z-boson    |Z   |Z   |23    |2     |MZ  |wZ   |1    |G  |Z         |Z
W-boson    |W+  |W-  |24    |2     |MW  |wW   |1    |G  |W^+       |W^-
Higgs      |h   |h   |25    |0     |Mh  |!wh  |1    |   |h         |h
electron   |e   |E   |11    |1     |Me  |0    |1    |   |e^-       |e^+
e-neutrino |ne  |Ne  |12    |1     |0   |0    |1    |L  |\nu_e     |\bar{\nu}_e
muon       |m   |M   |13    |1     |Mm  |0    |1    |   |\mu^-     |\mu^+
m-neutrino |nm  |Nm  |14    |1     |0   |0    |1    |L  |\nu_\mu   |\bar{\nu}_\mu
tau-lepton |l   |L   |15    |1     |Ml  |0    |1    |   |\tau^-    |\tau^-
t-neutrino |nl  |Nl  |16    |1     |0   |0    |1    |L  |\nu_\tau  |\bar{\nu}_\tau
d-quark    |d   |D   |1     |1     |0   |0    |3    |   |d         |\bar{d}
u-quark    |u   |U   |2     |1     |0   |0    |3    |   |u         |\bar{u}
s-quark    |s   |S   |3     |1     |0   |0    |3    |   |s         |\bar{s}
c-quark    |c   |C   |4     |1     |Mc  |0    |3    |   |c         |\bar{c}
b-quark    |b   |B   |5     |1     |Mb  |0    |3    |   |b         |\bar{b}
t-quark    |t   |T   |6     |1     |Mt  |wt   |3    |   |t         |\bar{t}
==================================================================================
\end{verbatim}
}
Each column is self-explanatory, and has information about 
particle name, symbol for particle, symbol for its anti-particle, PDG number, spin, mass, width, color charge, auxiliary field and latex expression.
All variables need to be defined either in ``{\tt Parameters}" or in ``{\tt Constraints}" except for a width 
that is prefixed with an exclamation mark ``!''.
For instance, the Higgs width is defined as ``!wh", which means it is calculated on the fly whenever needed.
\calchep supplies its value by running a separate decay process and the variable does not need to be defined in other model files such as 
``{\tt Parameters}" or ``{\tt Constraints}". \\

\item ``{\tt Parameters}" contains all necessary independent parameters that define a model and 
a numerical value is assigned to each parameter. 
     Users can change values of those parameters, when needed. Below is ``vars1.mdl'' for the Standard Model.
{\scriptsize 
\begin{verbatim}
 Parameters 
>Name       <| Value      |> Comment                                             <|
 alfEMZ      |0.0078180608|MS-BAR electromagnetic alpha(MZ)
 alfSMZ      |0.1172      |Srtong alpha(MZ) for running mass calculation
 GG          |1.238       |Running Strong coupling. The given value doesn't matter.
 SW          |0.481       |MS-BAR sine of the electroweak mixing angle
 Mm          |0.1057      |muon mass
 Ml          |1.777       |tau-lepton mass
 McMc        |1.2         |Mc(Mc)
 MbMb        |4.25        |Mb(Mb)
 Mtp         |175         |t-quark pole mass
 MZ          |91.187      |Z-boson mass
 Mh          |120         |higgs mass
 wt          |1.59        |t-quark width        (tree level 1->2x)
 wZ          |2.49444     |Z-boson width        (tree level 1->2x)
 wW          |2.08895     |W-boson width        (tree level 1->2x)
==================================================================================
\end{verbatim}
}
All other variables are calculated based on above parameters. \\

\item ``{\tt Constraints}" defines variables which depend on parameters or constraints that are defined previously. 
     Their values are calculated automatically when \calchep is running. ``func1.mdl'' is shown below.
{\scriptsize 
\begin{verbatim}
Standard Model
 Constraints 
>Name     <|> Expression                                               <|
EE         |sqrt(16*atan(1.)*alfEMZ)      % electromagnetic constant
CW         |sqrt(1-SW^ 2)                 % cos of the Weinberg angle
MW         |MZ*CW                         % W-boson  mass
c12        |sqrt(1-s12^ 2)                % parameter  of C-K-M matrix
c23        |sqrt(1-s23^ 2)                % parameter  of C-K-M matrix
c13        |sqrt(1-s13^ 2)                % parameter  of C-K-M matrix
Vud        |c12*c13                       % C-K-M matrix element
         .........................................
qcdOk      |initQCD(alfSMZ,McMc,MbMb,Mtp)
 Mb        |MbEff(Q)*one(qcdOk)
 Mt        |MtEff(Q)*one(qcdOk)
 Mc        |McEff(Q)*one(qcdOk)
=========================================================================
\end{verbatim}
}
Basic arithmetic operations such as +, -, /, *, \^ \; , Sqrt() are allowed.
Also external C functions are also allowed and they need to be defined in ``usrfun.c''.  \\

\item ``{\tt Vertices}" contains actual interactions of particles (``lgrng1.mdl''). 
Current version of \calchep allows interactions with 4 particles. 
An interaction is defined by a prefactor and Lorentz structure as shown below.
Interestingly, in the Lorentz part, ``/'' is not allowed and so any division should be carried out 
in the ``Factor'' column. 
The m2.m3 (m3.m4) represents the metric $g_{m_2 m_3}$, where m2 (m3) is the Lorentz index of the second (third) particle.
{\scriptsize 
\begin{verbatim}
Standard Model
 Vertices
A1   |A2   |A3   |A4   |>         Factor             <|>  Lorentz part                                                        <|
         .........................................
h    |W+   |W-   |     |EE*MW/SW                      |m2.m3
h    |Z    |Z    |     |EE/(SW*CW^ 2)*MW              |m2.m3
h    |h    |h    |     |-(3/2)*EE*Mh^ 2/(MW*SW)       |1
h    |h    |h    |h    |(-3/4)*(EE*Mh/(MW*SW))^ 2     |1
h    |h    |Z    |Z    | (1/2)*(EE/(SW*CW))^ 2        |m3.m4
h    |h    |W+   |W-   | (1/2)*(EE/SW)^ 2             |m3.m4
M    |m    |h    |     |-EE*Mm/(2*MW*SW)              |1
L    |l    |h    |     |-EE*Ml  /(2*MW*SW)            |1
C    |c    |h    |     |-EE*Mc/(2*MW*SW)              |1
B    |b    |h    |     |-EE*Mb/(2*MW*SW)              |1
T    |t    |h    |     |-EE*Mt  /(2*MW*SW)            |1
         .........................................
\end{verbatim}
}
The usual $i$ that appears in the Feynman rules is omitted by default. \\

\item ``{\tt Libraries}'' contains user-defined codes (``extlib1.mdl'' is shown as an example below.).
{\scriptsize 
\begin{verbatim}
Standard Model
Libraries
External libraries  and citation                                                 <|
   $CALCHEP/utile/usrfun.c
==================================================================================
\end{verbatim}
}

\end{enumerate}

%%%%%%%%%%%%%%%%%%%%%%%%%%%%%%%%%%%%%%%%%%%%%%%%%%
\subsection{Running examples}
%%%%%%%%%%%%%%%%%%%%%%%%%%%%%%%%%%%%%%%%%%%%%%%%%%

%%%%%%%%%%%%%%%%%%%%%%%%%%%%%%%%%%%%%%%%%%%%%%%%%%
\subsubsection{Tree level branching fractions of the Higgs}
%%%%%%%%%%%%%%%%%%%%%%%%%%%%%%%%%%%%%%%%%%%%%%%%%%

To warm up we start with an easiest example: two body decay of the Higgs in the Standard Model.
Now run \calchep and follow the steps below to open a numerical session.
\begin{enumerate}
\item Choose ``Standard Model'' in the main menu and turn on unitary gauge.
\item Select ``Enter Process''. \calchep shows the previous process on screen.
\item Input the two body decay of the Higgs as follows. We will not exclude any particles or any diagrams. So leave them as blanks.

\begin{verbatim}
Enter process: h -> 2*x
Exclude diagrams with
Exclude X-particles
\end{verbatim}
\item \calchep checks whether the directory ``results/" is empty. Users can either delete or rename results from the previous run.
\item In the next screen users can verify diagrams by choosing ``view diagrams''. Users can choose certain diagrams, if necessary.
\item To continue to a numerical session, select ``Squaring technique'' and then ``Make \& launch n\_calchep''
\item New GUI window will appear with different menus. 
\end{enumerate}

Once numerical session is opened, users can change parameters and impose cuts.
For detailed information regarding each menu, users should refer to the manual.
In this exercise, we simply use the last menu ``Easy 1\arrow 2". 
\calchep should show the total width with branching fractions on the next screen. 
For a different Higgs mass (or other parameters in general), users can play with ``parameter dependence'' in the menu. 
\calchep allows users to plot quantities, branching fractions and total width, 
for instance, as a function of one parameter in the model,  and output the data into a file. 

To get the total decay width, we can also run a script from the command line. 
Quit current numerical session by pressing {\tt Esc} key a few times. 
%Numerical session can be brought up again by simply running ``n\_calchep'' from ``results/'' directory. 
To use scripts, go to the ``results'' directory and run ``subproc\_cycle" as follows. 
{\small 
\begin{verbatim}
cd results
../bin/subproc_cycle 0
width(h)=0.00303927735
#Subprocess 2 ( h -> b, B ) width=0.0025155 Br=0.8276638524
#Subprocess 3 ( h -> c, C ) width=0.0001038 Br=0.03415285545
#Subprocess 4 ( h -> l, L ) width=0.0002501 Br=0.08228929815
#Subprocess 5 ( h -> m, M ) width=8.8605E-07 Br=0.0002915331172
#Subprocess 8 ( h -> A, Z ) width=9.049E-07 Br=0.0002977352495
#Subprocess 9 ( h -> A, A ) width=5.9964E-06 Br=0.001972969002
#Subprocess 10 ( h -> G, G ) width=0.00016209 Br=0.05333175664
\end{verbatim}
}
Here the input ``0'' after ``subproc\_cycle'' is the total number of events.
Output includes the total decay width as well as partial widths and branching fractions 
\footnote{The last two decay modes are added through an exercise in this note. 
If you haven't done so, you may not see them in the output.}. 
The numerical session can be called again anytime by running {\tt n\_calchep} which is in ``results'' directory.

%%%%%%%%%%%%%%%%%%%%%%%%%%%%%%%%%%%%%%%%%%%%%%%%%%
\subsubsection{4 body decay: $h \to  e^- + \bar{\nu}_e + \mu^+ + \nu_\mu$}
%%%%%%%%%%%%%%%%%%%%%%%%%%%%%%%%%%%%%%%%%%%%%%%%%%

Now let us look at a slightly more complicated exercise with the four body decay of the Higgs to two charged leptons via two $W$s.
We follow similar steps as before, but now input a different process
\begin{verbatim}
h -> e, Ne, M, nm
\end{verbatim}
There should be two diagrams, one with two $W$s and the other with one $W$.
The latter appears due to the Yukawa coupling of the muon, which is negligible.
Launch numerical session. Now the last menu in the previous run, ``Easy 1 \arrow 2'', does not appear in this example. 
To calculate the four body partial decay width, users need to run ``Vegas'', 
where the number of iterations and the number of MC points are set up. 
To calculate the width, simple choose ``Start integration'' and the result will be shown on screen.
To obtain a partial width for a different Higgs mass, 
users should provide a different numerical value to the mass variable in the menu ``Model parameters'' and run Vegas again.

\begin{figure}[t]
\centering
\epsfig{file=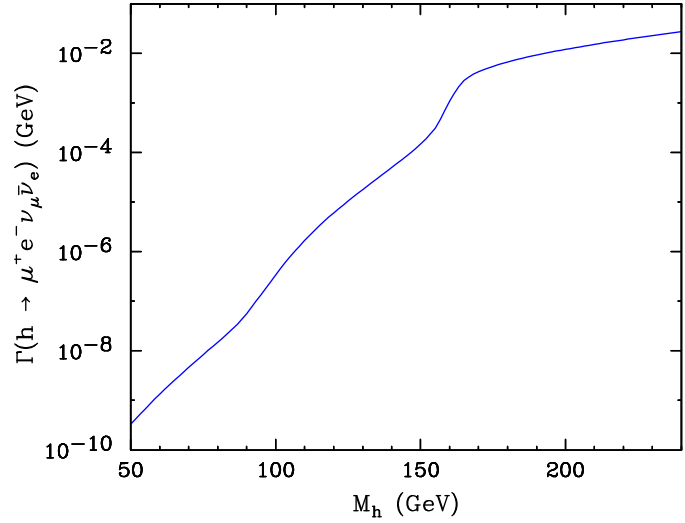,width=3.5in}
\caption{$\Gamma (h \to  e^- + \bar{\nu}_e + \mu^+ + \nu_\mu )$: the partial decay width of the Higgs to the $e^-\mu^+$ final state.}
\label{fig:htoee}
\end{figure}

Alternative way to compute it is to use a batch mode \footnote{
For the most recent version of the batch mode, see the manual. \cite{Belyaev:2012qa}.}. 
To use it, quit numerical session first and go to result directory and run ``name\_cycle'' that is in the ``bin'' directory.
\begin{verbatim}
../bin/name_cycle Mh 50 10 10
\end{verbatim}
This script computes the corresponding decay for 10 different Higgs masses starting from 50 GeV with 10 GeV interval.
Output is shown on screen and saved into a file as well. 
\begin{verbatim}
Mh=50 sigma=3.2620E-10[pb]
Mh=60 sigma=1.3323E-09[pb]
Mh=70 sigma=4.6880E-09[pb]
Mh=80 sigma=1.4999E-08[pb]
Mh=90 sigma=5.6488E-08[pb]
Mh=100 sigma=3.4003E-07[pb]
Mh=110 sigma=1.6884E-06[pb]
Mh=120 sigma=6.2347E-06[pb]
Mh=130 sigma=1.8358E-05[pb]
Mh=140 sigma=5.0125E-05[pb]
See Mh_tab_2_11 file.
\end{verbatim}
Taking the data file with more points, one can plot the four body decay width of the Higgs in the 
$e^-\mu^+$ final state, which is shown in Fig. \ref{fig:htoee} \footnote{Users are supposed to use their own plotting package for this.}. 
One can notice the slight change in the slopes near $M_W$ and $2 M_W$. 
The former is due to the fact that one of the $W$ becomes onshell, 
while in the latter both $W$ are onshell.

%

%%%%%%%%%%%%%%%%%%%%%%%%%%%%%%%%%%%%%%%%%%%%%%%%%%
\subsubsection{$p \bar{p} \to W b \bar{b}$}
%%%%%%%%%%%%%%%%%%%%%%%%%%%%%%%%%%%%%%%%%%%%%%%%%%

In this example, we consider the scattering process $p \bar{p} \to W b \bar{b}$ at the Tevatron. 
Steps are given as follows.
\begin{enumerate}
\item Open numerical session with the following process. 
We will call a proton as ``p'' and will include the first generation quarks and gluon in it.
\begin{verbatim}
Enter process:  p, p -> W+, b, B
composite `p' consists of:  u, d, U, D, G
Exclude diagrams with
\end{verbatim}
\item In ``IN State'', choose PDF sets and set up the momentum of each beam. 
For instance, CTEQ5M(proton) for ``S.F.1'' and CTEQ5M(anti-proton) ``S.F.2'', and 980 GeV for each beam.
\item Cuts, kinematics and regularizations are useful for fast and more accurate computation. An example for cuts is shown below.
\begin{verbatim}
 #Cuts 
*** Table ***
 Cuts  
!|  Parameter  |> Min bound <|> Max bound <|
 |T(b)         |20           |             |
 |T(B)         |20           |             |
 |N(b)         |-5           |5            |
 |N(B)         |-5           |5            |
 |J(b,B)       |0.4          |             |
============================================
\end{verbatim}

There are many other available variables that are used as cuts (get help by pressing {\tt F1}.): 
A (angle in degree units), 
C (cosine of angle), 
J  (jet cone angle, which is defined as $\sqrt{\Delta y^2 + \Delta \phi^2}$, 
where $\Delta y$ is the pseudo-rapidity difference and $\Delta \phi$ is the azimuthal angle 
difference between two particles), 
E (energy of the particle set), 
M (mass of the particle set), 
P (cosine in the rest frame of pair), 
T (transverse momentum of the particle set), 
S (squared invariant mass of the particle set), 
Y (rapidity of the particle set), and 
U (user's implemented function). The character string following U is
passed on to the user C function usrfun(str) which, as it is
assumed, calculates a corresponding value. See Section 2.3 for further
explanations. \\

To set up relevant kinematics and regularizations, users need to take a look at diagrams and specify them.
For instance, in this example, $b$ and $\bar{b}$ may be combined.
\begin{verbatim}
#Kinematical_scheme 
12 -> 45 , 3
45 -> 4 , 5
\end{verbatim}
The corresponding numbers are from the name of the process in the top line of numerical session of \calchep.
They are numbered consecutively.
As one can see from the process label,
\begin{verbatim}
(sub)Process: u, D -> W+, b, B
\end{verbatim}
 \calchep assigns 3 for {\tt W+}, 4 for {\tt b} and 5 for {\tt B}. 
 
In the regularizations users specify a list of dangerous propagators. \calchep performs regularizations of the squared matrix element according to the contents of this list.
\begin{verbatim}
#Regularization 
*** Table ***
 Regularization 
 Momentum    |> Mass  <|> Width <| Power|
45           |MZ       |wZ       |2     |
45           |Mh       |wh       |2     |
34           |Mt       |wt       |2     |
35           |Mt       |wt       |2     |
=========================================
\end{verbatim}

To see distributions in \calchep, they need to be set up before VEGAS is called. 
Examples are list below. 
\begin{verbatim}
Parameter_1  |>   Min_1   <|>   Max_1  <|
   M(W+, b)         80           200
   M(W+, B)         80           200   
   M(b,B)            0           200
=========================================
\end{verbatim}
Once users set up distributions, click on ``Start integration'' to get the total cross sections.
For plots, select ``Display Distributions''. 
To improve errors and {\tt chi**2}, increase {\tt nSess\_1} and {\tt nCalls\_1} (and  {\tt nSess\_2} and {\tt nCalls\_2} for 2 dimensional plots). 

\end{enumerate}

%%%%%%%%%%%%%%%%%%%%%%%%%%%%%%%%%%%%%%%%%%%%%%%%%%
\subsubsection{$t \bar{t}$ production cross section at the Tevatron}
%%%%%%%%%%%%%%%%%%%%%%%%%%%%%%%%%%%%%%%%%%%%%%%%%%
Steps for $p \bar{p} \to t \bar{t}$ are summarized as follows.
\begin{enumerate}
\item Run $p \bar{p} \to t \bar{t}$ similarly with minor change as follows.
\begin{verbatim}
Enter process:  p, p -> t, T
composite `p' consists of:  u, d, s, U, D, S, G
Exclude diagrams with  A, W+, W-, Z, h
\end{verbatim}
The last step forces \calchep to include diagrams which are mediated by strong interaction only.

\item Open numerical session and turn on PDF with CTEQ6L (for proton and anti-proton) and set the momentum of beams. 
This time we set QCD scale to {\tt Mt/2} ($m_{top}/2$), which is known to minimize the difference between leading order and next-leading order cross sections.
All this information is saved in files, ``prt\_\#'' and ``session.dat".
\begin{verbatim}
#Initial_state  inP1=9.800000E+02  inP2=9.800000E+02
 Polarizations= { 0.000000E+00  0.000000E+00 }
  StrFun1="PDT:cteq6l(proton)" 2212
  StrFun2="PDT:cteq6l(anti-proton)" -2212
\end{verbatim}
and
\begin{verbatim}
#QCD alphaPDF=1 alpha(MZ)=1.172000E-01 NF=5 Order=2 
     MbMb=4.200000E+00 Mtp=1.750000E+02 Scale= Mt/2
\end{verbatim}

\item  Use ``subproc\_cycle" to get the total cross section.

{\scriptsize
\begin{verbatim}
../bin/subproc_cycle 10
#Subprocess 1 ( u, U -> t, T ) Cross section = 6.1934E+00, 61934 events
#Subprocess 2 ( d, D -> t, T ) Cross section = 1.1791E+00, 11791 events
#Subprocess 3 ( s, S -> t, T ) Cross section = 4.1461E-03, 41 events
#Subprocess 4 ( U, u -> t, T ) Cross section = 1.3220E-02, 132 events
#Subprocess 5 ( D, d -> t, T ) Cross section = 1.9288E-02, 192 events
#Subprocess 6 ( S, s -> t, T ) Cross section = 4.1498E-03, 41 events
#Subprocess 7 ( G, G -> t, T ) Cross section = 4.2567E-01, 4256 events
Sum of distributions is stored in file distr_1_7
Total Cross Section 7.8389739 [pb]
see details in prt_1 - prt_7 files
\end{verbatim}
}
Total cross section summed over all subprocesses is reported at the end of output.
The parameter that follows ``subproc\_cycle'' is mandatory and is the total integrated luminosity in [1/fb] unit.
Events are not generated with ``subproc\_cycle 0''. 

\item Now we can calculate the top mass dependence in the total cross sections using ``name\_cycle\_plus''
\footnote{This script is written by A. Pukhov but is not included in the \calchep package. Users need to download from 
http://susy.phsx.ku.edu/$\sim$kckong/tasi/name\_cycle\_plus.}.
\begin{verbatim}
../bin/name_cycle_plus Mtp 165 1 11
Mtp=165 sigma=9.6448496[pb]
Mtp=166 sigma=9.3545769[pb]
Mtp=167 sigma=9.0866098[pb]
Mtp=168 sigma=8.8071721[pb]
Mtp=169 sigma=8.5578445[pb]
Mtp=170 sigma=8.3177445[pb]
Mtp=171 sigma=8.064445[pb]
Mtp=172 sigma=7.843451[pb]
Mtp=173 sigma=7.6188025[pb]
Mtp=174 sigma=7.3939838[pb]
Mtp=175 sigma=7.1815069[pb]
See Mtp_cycle file.
\end{verbatim}
As expected from above output, ``name\_cycle\_plus'' reports the production cross section for 11 different masses with 1 GeV interval, 
starting from 165 GeV.

\item  If distributions are defined before ``subproc\_cycle'' is called, \calchep generates the corresponding distributions for each subprocess.
Users can see them simply typing
\begin{verbatim}
show_distr distr_#
\end{verbatim}
To combine distributions for all subprocess, run 
\begin{verbatim}
sum_distr distr_1 distr_2 > distr_sum
\end{verbatim}
and run ``show\_distr distr\_sum" for view.

\item ``event\_mixer'' combines generated events files and write output in LHE format, that can be then used in any other event generators. 
It needs two inputs: the number of events and a directory where events files exist.
If current path is `results' directory, simply type
\begin{verbatim}
../bin/event_mixer 10000 .
total cross section 7.175E+00
Max number of events 71503
\end{verbatim}
and this combines all events in 7 subprocess and weight them according to their individual cross sections.
The default output file is ``event\_mixer.lhe'' and users can generate simple distributions with this LHE file, using ``lhe2tab''.
For example, try the following exercise.
{\small 
\begin{verbatim}
../bin/lhe2tab "M(6,-6)" 300 1000 100 < event_mixer.lhe > Mtt.txt
\end{verbatim}
}
The generated ``Mtt.txt'' files contains differential cross section as a function of the invariant mass of top (6) and anti-top (-6).

\end{enumerate}

%    
 %   - u, U -> t, T -> W+, b, W-, B -> e, Ne, M, nm, b, B
%    - various distributions: M(e,b), M(e,B), M(e,Ne,B), M12, M(e,Ne) etc
 %   - try to understand invariant mass distributions.
%  SUSY \cite{Semenov:2002xn}

%%%%%%%%%%%%%%%%%%%%%%%%%%%%%%%%%%%%%%%%%%%%%%%%%%
\subsubsection{User defined variables, and constraints in external C file}
%%%%%%%%%%%%%%%%%%%%%%%%%%%%%%%%%%%%%%%%%%%%%%%%%%

User programs in \calchep provide users with possibility to attach his/her own codes to the {\tt n\_calchep}. 
In this way users are able to expand the set of phase space functions for cuts and histograms and 
implement new structure functions. We will discuss three examples: transverse mass, $M_{T2}$ and the Higgs effective coupling.
We define the first two quantities in ``usrfun.c'' that is found under ``utile'' directory. 
An example is included as shown below.
{\small
\begin{verbatim}
double usrfun(char * name, double * pvect)
{   
  if(strcmp(name,"MT")==0)  {    /* Transverse mass  */

    double tmass, ET1, ET2;
    double pp1[4];
    double pp2[4];
    int    k;
    
    for(k=0; k<4; k++)
      {
	          pp1[k] = pvect[4*(3-1)+k];
	          pp2[k] = pvect[4*(4-1)+k];
      }
    
    ET1 = sqrt( pow(pp1[1],2) + pow(pp1[2],2) );
    ET2 = sqrt( pow(pp2[1],2) + pow(pp2[2],2) );
    
    tmass = 2 * ( ET1 * ET2 - pp1[1] * pp2[1] - pp1[2] * pp2[2] ); 
    return sqrt(tmass);
  }

  if(strcmp(name,"MT2")==0)  {    /*  MT2 for massless particles  */
    double tmass2,ET1, ET2;
    double pp1[4];
    double pp2[4];
    int    k;
    
    for(k=0; k<4; k++)
      {
	          pp1[k] = pvect[4*(3-1)+k];
	          pp2[k] = pvect[4*(5-1)+k];
      }
    
    ET1 = sqrt( pow(pp1[1],2) + pow(pp1[2],2) );
    ET2 = sqrt( pow(pp2[1],2) + pow(pp2[2],2) );
    
    tmass2 = 2 * ( ET1 * ET2 + pp1[1] * pp2[1] + pp1[2] * pp2[2] ); 

    return sqrt(tmass2);
  }

  /* original USRFUNC  */

   fprintf(stdout," usrfun(char* name)  called with parameter %s\n"
                  " But is not  defined!\n",name);
   sortie(54);
   return 0.;
}
\end{verbatim}
}
Users also need to specify its location in a model file, ``extlib9.mdl''.
\begin{verbatim}
Standard Model
Libraries
External libraries  and citation                             <|
   $CALCHEP/utile/usrfun.c
==============================================================
\end{verbatim}

Now let us consider single production of the $W$ and its leptonic decay.
A well known and useful quantity in this process is the transverse mass, 
which is not implemented in \calchep.

\begin{enumerate}
\item Run p, p \arrow  e, Ne at the Tevatron (or at the LHC).
\item To take a look at the invariant mass and the transverse mass of $e$ and $\bar{\nu}_e$, we defined distributions as follows.
\begin{verbatim}
Distributions
Parameter_1|>    Min_1    <|>    Max_1    <|
M(e,Ne)    |             0 |           100 |           
UMT        |             0 |           100 |          
============================================
\end{verbatim}
Here new distribution ``UMT'' is one of the functions that we defined in ``usrfun.c'' above.
For \calchep to understand that this is a user defined function, 
the letter ``U'' should come with the name of user variables. 
This can be used in ``cuts'' as well.
Unfortunately \calchep currently does not support having the particle names inside user-defined functions.
\end{enumerate}

Another good example of user-defined routines is the Higgs effective couplings to two photons.
It is well known that the dominant production of the Higgs at the LHC is via  glue-glue fusion.
The main contribution arises at loops and unfortunately many tree-level event generators including \calchep do not treat this interaction.

However one can parameterize them in term of higher dimensional operators and often the coefficients of such operators 
involve complicated expressions such as integrals. In this case, one can perform calculation in a separate code, ``usrfun.c'' and return 
the result into \calchep. We will come back to this later in section \ref{sec:higgseffective}. 
For discussion in the rest of this section, suppose we already have this interaction implemented.

Next example is single production of the Higgs and its decay to two leptons via two $W$s.
\begin{enumerate}
\item To Run $g g \to h \to W^+ W^- \to e^- \bar{\nu}_e \,  \mu^+ \nu_\mu$, type
\begin{verbatim}
Enter process: G, G -> e, Ne, M, nm
Exclude diagrams with m
\end{verbatim}
\item Kinematics, regularizations and distributions are set up as follows.

{
\begin{verbatim}
#Kinematical_scheme 
12 -> 34 , 56
34 -> 3 , 4
56 -> 5 , 6

#Regularization 
 Regularization 
 Momentum    |> Mass  <|> Width <| Power|
34           |MW       |wW       |2     |
56           |MW       |wW       |2     |
3456         |Mh       |wh       |2     |
=========================================

#Distributions
Parameter_1|>    Min_1    <|>    Max_1    <|
M(e,Ne)    |   0.000000E+00|   1.000000E+02|          
M(M,nm)    |   0.000000E+00|   1.000000E+02|    
M12        |   0.000000E+00|   2.000000E+02|        
UMT2       |   0.000000E+00|   1.000000E+02|        
============================================
\end{verbatim}
}
The ``M12'' distribution is the invariant mass of the first two particles, i.e., G and G. Therefore this is equivalent to $\sqrt{\hat s}$ and 
the invariant mass of all particles in the final state, which should peak at the Higgs mass in this example. 
The ``UMT2'' is the $M_{T2}$ distribution taking $e^-$ and $\mu^+$ as visible particles and the two neutrinos as the missing particles. 
Once they are set up, perform Vegas integration, which will report cross sections at the end of the run.
To see previously-defined distributions, click on ``Display Distributions''. 
\end{enumerate}

%%%%%%%%%%%%%%%%%%%%%%%%%%%%%%%%%%%%%%%%%%%%%%%%%%
\subsubsection{KK photon annihilation in mathematica}
%%%%%%%%%%%%%%%%%%%%%%%%%%%%%%%%%%%%%%%%%%%%%%%%%%

One of excellent features of \calchep is that it provides analytic expressions for squared matrix elements 
in FORM, REDUCE and Mathematica. To learn how to exploit this analytic feature, we consider 
KK photon annihilation in Universal Extra Dimensions (UED). We will follow the well known procedure in literature \cite{Servant:2002aq}.

The relic abundance of dark matter $\chi$ is found by solving the Boltzmann equation 
for the evolution of the $\chi$ number density $n$
\begin{equation}
\frac{d n}{ d t} = -3 Hn - \langle \sigma v \rangle ( n^2 - n^2_{eq})\ ,
\end{equation}
where $H$ is the Hubble parameter, $v$ is the relative velocity between two $\chi$'s, 
$\langle \sigma v \rangle$ is the thermally 
averaged total annihilation cross-section times relative velocity, 
and $n_{eq}$ is the equilibrium number density. 
$\langle \sigma v \rangle$ is often approximated by the non-relativistic expansion
\begin{equation}
\langle \sigma v \rangle = a + b \langle v^2 \rangle + {\cal O}(\langle v^4 \rangle) 
\approx a+ 6 b /x + {\cal O}\left( \frac{1}{x^2}\right) \ ,
\end{equation}
where $x = \frac{m}{T}$ is the ratio of the dark matter mass to the temperature. 
By solving the Boltzmann equation analytically with appropriate 
approximations \cite{Servant:2002aq}, 
the abundance of $\chi$ is given by
\begin{equation}
\Omega_\chi h^2 \approx \frac{1.04 \times 10^9}{M_{Pl}}\frac{x_F}{\sqrt{g_\ast(x_F})} \frac{1}{a+3 b/x_F }\ ,
\end{equation}
where the Planck mass $M_{Pl} = 1.22\times 10^{19}$ GeV and
$g_\ast$ is the total number of effectively massless degrees of freedom. 
The freeze-out temperature, $x_F$, is found iteratively from 
\begin{equation}
x_F = \ln \left ( c(c+2) \sqrt{\frac{45}{8}} \frac{g}{2\pi^3} \frac{m M_{Pl} (a+6b/x_F)}{\sqrt{g_\ast(x_F) x_F}}  \right )\ ,
\end{equation}
where the constant $c$ is determined  empirically by comparing to numerical
solutions of the Boltzmann equation. 
At the end the calculation of the relic abundance becomes computation of annihilation cross sections of relevant processes.

In the case of Minimal UED (MUED), KK photon is a good dark matter candidate and a pair of KK photons annihilates to the SM fermions and Higgses.
Here are the steps to get annihilation cross sections for \calchep.
\begin{enumerate}
\item Download models files for the dark matter model, in this case MUED
\footnote{The model files can be obtained from the web site for this TASI tutorial, or directly from 
http://susy.phsx.ku.edu/$\sim$kckong/tasi/MUED.tar \cite{Datta:2010us}.  
It is important to check the model file has no errors with \calchep version that users are running. 
This can be done by simply selecting ``CHECK MODEL'' under ``Edit model'' in the main menu.}.
\item Unpack: {\tt tar -xvf MUED.tar} and import this model by selecting ``IMPORT of MODELS'' in the main \calchep menu 
(One can simply copy files into {\tt models} under user's working directory but the model number needs to be changed consecutively.).
\item Run $B1 ~B1 \to ~e~ E$ for a typical KK photon annihilation. Users should see 4 diagrams with two kinds of KK fermions.
\item After squaring diagrams, go to ``Symbolic calculations'' and save the squared matrix elements in Mathematica code. 
	Users should see that ``symb1.m'' is generated under ``results'' directory.
\item Now in Mathematica, run ``sum\_int.m''  which is found in a directory called ``utile'' (or slightly modified version can be downloaded from 

http://susy.phsx.ku.edu/$\sim$kckong/tasi/TASI\_MUED\_B1B1\_ee.nb.).
\item After loading the package, run the following command in Mathematica.

\begin{verbatim}
sum = 0
addToSum:= sum 
         = sum + Simplify[  totFactor numerator/denominator 
          /. substitutions ];
\end{verbatim}

\item Then load the generated matrix element in ``symb1.m''. By simply typing ``sum'' we obtain the squared matrix element, $| {\cal M}|^2$. 
``symb1.m'' includes squared matrix elements as well as the corresponding diagrams. 
Each squared diagram is computed by multiplying three quantities: totFactor, numerator and denominator. 

\item Now the corresponding annihilation cross section is obtained by integrating over phase space.
\begin{equation}
\sigma = -  \frac{1}{16 \pi \lambda_{12}} \int_{t_0}^{t_\pi} d t  \, | {\cal M}|^2 \, ,
\end{equation}
where $\lambda_{ij} = \lambda(s, m_i^2, m_j^2) = ( s- m_i^2 -m_j^2)^2 - 4 m_i^2 m_j^2 $ for a particle scattering, $1,2 \to 3, 4$.
The lower (upper) bound is given by 
$t_0 (t_\pi) = \frac{1}{4 s} \left ( (m_1^2 - m_2^2-m_3^2 + m_4^2)^2 - ( \sqrt{\lambda_{12}} \mp \sqrt{ \lambda_{34}}) \right )^2 $. 
Here $s$, $t$ and $u$ are the Mandelstam variables and $m_1 = m = m_2$ and $m_3 = 0 =m_4$ in this example.

\item By simplifying the cross section, we should obtain the following expression
\begin{equation}
\sigma (B1 B1 \to e^+ e^-) = \frac{g_1^4 (Y_{e_L}^4 + Y_{e_R}^4)}{72 \pi s^2 \beta^2} 
\left (     10 ( 2 m^2 + s) tanh^{-1} \beta - 7 \beta s   \right ) \, ,
\end{equation}
where $\beta = \sqrt{ 1 - \frac{4 m^2}{s}}$, $Y$'s are the Hypercharges of electrons and $g_1$ is the strength of the $U(1)_Y$ interaction. 
\item Finally two leading terms (a and b terms) are obtained by expanding in terms of the relative velocity. 
\end{enumerate}

%%%%%%%%%%%%%%%%%%%%%%%%%%%%%%%%%%%%%%%%%%%%%%%%%%
\subsection{Implementing new particles and new interactions}
%%%%%%%%%%%%%%%%%%%%%%%%%%%%%%%%%%%%%%%%%%%%%%%%%%

%%%%%%%%%%%%%%%%%%%%%%%%%%%%%%%%%%%%%%%%%%%%%%%%%%
\subsubsection{Higgs effective couplings}
\label{sec:higgseffective}
%%%%%%%%%%%%%%%%%%%%%%%%%%%%%%%%%%%%%%%%%%%%%%%%%%

For the Higgs effective couplings to photons and gluons, we do not need to include new particles but 
need to include new interactions of the Higgs (new in a sense that \calchep does not have it). 
For a relatively light Higgs, these may be expressed in terms of dimension 5 operators, 
\begin{eqnarray}
{\cal L}_{ggh} &=& -\frac{1}{4} g_{ggh} G_{\mu\nu}^a G^{\mu\nu a} h  \, ,\\
{\cal L}_{\gamma\gamma h} &=& -\frac{1}{4} g_{\gamma \gamma h} F_{\mu\nu} F^{\mu\nu} h \, ,
\end{eqnarray}
and the couplings are calculated approximately as
\begin{eqnarray}
g_{ggh}  &=& \frac{\alpha_S}{3\pi v} \left ( 1 + \frac{7}{30} + + \frac{2}{21} \tau_t^2 +  \frac{26}{525} \tau_t^3  + \cdots  \right ) \, , \\
g_{\gamma \gamma h} &=& - \frac{\alpha}{\pi v} \frac{47}{18} \left (  1 + 
\frac{66}{235} \tau_w + \frac{228}{1645} \tau_w^2 + \frac{696}{8225} \tau_w^3 + \frac{5248}{90475} \tau_w^4 \right. \nonumber \\
&& \hspace*{1cm} + \frac{1280}{29939} \tau_w^5 + \frac{54528}{1646645} \tau_w^6 
 \left . -  \frac{56}{705} \tau_t -  \frac{32}{987} \tau_t^2 + \cdots    \right ) \, , 
\end{eqnarray}
where $\tau_t =  \frac{m_h^2}{4 m_t^2} $ and $\tau_w =  \frac{m_h^2}{4 m_W^2}$
\footnote{MadGraph\cite{Alwall:2011uj} has the same implementation. See the following webpage for more information: https://server06.fynu.ucl.ac.be/projects/madgraph/wiki/Models/HiggsEffective.}.
One can compute Feynman rules for these interaction easily and implement them in ``Vertices'' in \calchep 
\footnote{For a different approach, users are encouraged to investigate 
FeynRules \cite{Christensen:2008py} and LanHEP \cite{Semenov:2010qt}.}.
{\scriptsize
\begin{verbatim}
Standard Model
 Vertices
A1   |A2   |A3   |A4   |>         Factor             <|>  Lorentz part            <|
===================================================================================
G    |G    |h    |     |(GG^2/(4*pi))/(12*pi*vev)*(-4)|p1.p2*m1.m2 - m2.p1*m1.p2
A    |A    |h    |     |chAA                          |p1.p2*m1.m2 - m2.p1*m1.p2
===================================================================================
\end{verbatim}
}
We only take the leading term in the couplings for simplicity.
The structure of this file is very easy to understand, which is again big advantage of using \calchep.
In the ``Lorentz'' part, ``p1'' denotes the momentum of the first particle in the particle listing, A1, which is G, 
``p2'' is the momentum of the second particle, A2, which is G again in this case.
``m1'' and ``m2'' are Lorentz indices and hence p1.m2 (or m2.p1) means the momentum of the first particle A1, 
which carries the Lorentz index of the second particle A2.
The m1.m2 is the metric, $g_{m1 m2}$. 
In the ``Factor'' we can include full expression of the coupling as we have done for $g_{ggh}$ or 
we can define the coupling in the ``Constraints'', 
 \begin{verbatim}
    Standard Model
 Constraints 
>Name     <|> Expression                                 <|
==========================================================
chAA       |hAA(EE,vev)
==========================================================
\end{verbatim}
and include definition of hAA(EE,vev) in the ``usrfun.c''.
\begin{verbatim}
double hAA(double EEE, double VVV)
{
  double hAAcoupling;
  double alpha;
  double pi;

  pi=acos(-1);
  alpha = EEE*EEE/(4*pi);

  hAAcoupling = alpha/(pi*VVV) * 47/18;  
   /*  first term from the effective coupling of 
                        the higgs to two photons*/

    return hAAcoupling;
}
\end{verbatim}
One of the mode files  ``extlib\#.mdl'' has path to this user-defined code.
\begin{verbatim}
SM with Gprime
Libraries
External libraries  and citation                          <|
   $CALCHEP/utile/usrfun.c
============================================================
\end{verbatim}
%

%%%%%%%%%%%%%%%%%%%%%%%%%%%%%%%%%%%%%%%%%%%%%%%%%%
\subsubsection{Implementing a color-octet vector boson}
\label{sec:Gprime}
%%%%%%%%%%%%%%%%%%%%%%%%%%%%%%%%%%%%%%%%%%%%%%%%%%

A new particle can be implemented easily in \calchep using GUI interface.
Suppose we are interested in a model with one new particle, color-octet vector boson $G^\prime$, 
which has interaction with the SM quarks ($q$ and $\bar q$), 
\begin{equation}
{\cal L} = g_3 \bar{q} \left ( q_V + q_A \gamma_5  \right ) \gamma^\mu \frac{\lambda^a}{2} q G^{\prime a}_\mu \, ,
\end{equation}
where $\lambda^a$ is the Gell-Mann matrix, $g_3$ is the coupling strength of QCD and $\gamma_\mu$s are
the Dirac $\gamma$ matrices.  
\begin{enumerate}

\item First we define the particle in ``Particles''.
{\scriptsize
    \begin{verbatim}
SM with Gprime
 Particles 
 Full  name |>A <|>A <|number |2*spin|mass|width|color|aux|>LaTex(A)<|>LaTeX(A+)<|
gluon       |G   |G   |21     |2     |0   |0    |8    |G  |g         |g
                 ...................................................
t-quark     |t   |T   |6      |1     |Mt  |wt   |3    |   |t         |\bar{t}
Gprime      |~G  |~G  |3100021|2     |MGP |!wGP |8    |G  |          |
==================================================================================
 \end{verbatim}
 }
 Most of necessary inputs are easy to understand. 
 The exclamation mark ``!'' in the particle width means \calchep will run the decay process and calculate the width on the fly. 
 The newly introduced variable, MGP is the mass of the new particle and we will define it in ``Parameter''. \\

  \item For the interactions, \calchep already knows about the Gell-Mann matrices from the quantum charges 
  and users do not need to include them in the vertices. 
   {\scriptsize
    \begin{verbatim}
SM with Gprime
 Vertices
A1   |A2   |A3   |A4   |>         Factor             <|>  Lorentz part                                                        <|
D    |d    |~G   |     |GG                            |G(m3)*(qV + qA * G5)
U    |u    |~G   |     |GG                            |G(m3)*(qV + qA * G5)
T    |t    |~G   |     |GG                            |G(m3)*(tV + tA * G5)
           ...................................................
    \end{verbatim}
} 
The ``G'' and ``G5'' denote the Dirac $\gamma$ matrices and the ``GG'' is the 
  strong coupling constant. Since we have introduced new variables (qV, qA, tV, tA and MGP), we should define them 
  either in ``Parameters'' or in  ``Constraints''. We do this in the ``Parameters'' for this exercise. 
   {\scriptsize
 \begin{verbatim}
SM with Gprime
 Parameters 
>Name         <| Value       |> Comment                                                 <|
 alfEMZ        |0.0078180608 |MS-BAR electromagnetic alpha(MZ)
               ...................................................
qA             |1            |  axial coupling of Gprime to qqbar
qV             |0            |  vector coupling of Gprime to qqbar
tA             |-1           |  axial coupling of Gprime to ttbar
tV             |0            |  vector coupling of Gprime to ttbar
MGP            |1000         |  mass of Gprime
==================================================================
   \end{verbatim}
}
This completes implementation of a color-octet vector boson and its interaction to the SM quarks.

 \end{enumerate}

%%%%%%%%%%%%%%%%%%%%%%%%%%%%%%%%%%%%%%%%%%%%%%%%%%
\section{PYTHIA}
%%%%%%%%%%%%%%%%%%%%%%%%%%%%%%%%%%%%%%%%%%%%%%%%%%

\pythia is frequently used for event generation in high-energy physics. The
emphasis is on multi-particle production in collisions between elementary particles. This
in particular means hard interactions in $e^+e^-$, $p p$ and $e p$ colliders, although also other
applications are envisaged. The program is intended to generate complete events, in as
much detail as experimentally observable ones, within the bounds of our current understanding
of the underlying physics. 

In this tutorial, we will consider a few specific examples.
Unlike \calchep, \pythia requires a main driver, which will be compiled together with \pythia code. 
We will use Fortran as our main compiler\footnote{
We will assume that users are familiar with Fortran.
The most recent version is \pythia 8 and is written in C++.}.

%%%%%%%%%%%%%%%%%%%%%%%%%%%%%%%%%%%%%%%%%%%%%%%%%%
\subsection{Installation}
%%%%%%%%%%%%%%%%%%%%%%%%%%%%%%%%%%%%%%%%%%%%%%%%%%

Make sure you have a Fortran compiler, e.g. g77, gfortran and etc. 
We will use \pythia version 6.4.25 and it can be downloaded from {\tt http://home.thep.lu.se/$\sim$torbjorn/Pythia.html}.  
Create ``pythia-6.4.25.o" by typing ``g77 -c pythia-6.4.25.f".
You should see pythia-6.4.25.o if the compile was successful.
If you have a different compiler, use that one instead.
Download an example Fortran code (``main61.f'') for the most trivial test, and copy it to the same directory where you have \pythia.
Compile it with \pythia and make an executable, by typing ``g77 -o main61.x main61.f pythia-6.4.25.o".
Run ``./main61.x'' in your terminal and take a look at output on your screen.

To make compile processes easier, often ``makefile'' is used. An example is shown below. 
 \begin{verbatim}
# Makefile for pythia examples

 FF = gfortran
# FF = g77 
# this is ok with gfortran from fink
FFLAGS = -g -static -w -fno-second-underscore   

# For MAC: use the following FFLAGS     
# FFLAGS = -gdwarf-2 -static -w -fno-second-underscore

OBJS = main61.o ../pythia-6.4.25.o
EXEC = pythia.x
#LIB  =  /cern/pro/lib/libmathlib.a 

all: $(EXEC)

$(EXEC): $(OBJS)
        $(FF) -o $(EXEC) $(OBJS) $(LIB)

.f.o:
        $(FF) $(FFLAGS) -c $<
 \end{verbatim}
In this example of makefile, the executable is named as ``pythia.x'' and gfortran is used as a Fortran compiler.
If additional libraries are needed, one can add them to the ``LIB'' variable and uncomment out the corresponding line.

%%%%%%%%%%%%%%%%%%%%%%%%%%%%%%%%%%%%%%%%%%%%%%%%%%
\subsection{Running examples}
%%%%%%%%%%%%%%%%%%%%%%%%%%%%%%%%%%%%%%%%%%%%%%%%%%

%%%%%%%%%%%%%%%%%%%%%%%%%%%%%%%%%%%%%%%%%%%%%%%%%%
\subsubsection{$t\bar{t}$ production}
%%%%%%%%%%%%%%%%%%%%%%%%%%%%%%%%%%%%%%%%%%%%%%%%%%

The beginning of \pythia main driver include many lines of common blocks and 
we will not go over their details. Instead we focus on specific examples.
The first example with \pythia is $t\bar{t}$ production at the Tevatron. 
Here we show general structure of the main code\footnote{The code below is not complete and users 
will not be able to compile with this. Part of the code is shown for discussion. 
Complete source can be found from http://susy.phsx.ku.edu/$\sim$kckong/tasi/. 
It also contains many comments. 
They do not affect the running of the code and 
are added for user's convenience.}. It is readable, if users understand basics of collider physics. 

{\small
 \begin{verbatim}
C-------------------------------------------
C...First section: initialization.
C-------------------------------------------

C...Number of events
      nevpythia=10000

C...MSEL selects individual process      
       MSEL=6     ! t quark,

C...Sample code to force only decays of interest
      do i=190,208  ! Turn off all W decays
        mdme(i,1) = 0   ! See the manual for the meaning of mdme 
      enddo
      mdme(206,1) = 1  ! Turn on electron + neutrino

C particles masses 
      PMAS(6,1)=172.            ! top mass
      PMAS(24,1)=80.            ! W mass

C...If interested only in cross sections and resonance decays: 
C...switch off initial and final state radiation, 
C...multiple interactions and hadronization.
      MSTP(61)=0 ! initial state radiation [0] off, [D=2] on
                 ! [1]: on for QCD radiation in hadronic events 
                 !       and QED radiation in leptonic ones
                 ! [2]: on for QCD/QED radiation in hadronic events 
                 !         and QED radiation in leptonic ones
      MSTP(71)=0 ! final state radiation [0] off, [D=1] on
      MSTP(81)=0 ! multiple interactions [0] off, [D=1] on
      MSTP(111)=0! fragmentation and decay [0] off, [D=1] on
      MSTP(91)=0 ! No primordial kT

C... PDF: [D=8], page 200
c      MSTP(32) = 4  ! Q^2 value set, root-s-hat
c      MSTP(32) = 11 ! Q^2 = (m3+m4)^2/4
c      MSTP(52)=2    ! to use external PDF lib, [D=1]
c      MSTP(51)=7    ! [D=7]     
                     ! choose pdf. 7:CTEQ5L LO, 8: CTEQ5M1

C...Initialization for the Tevatron or LHC or ILC. 
C      CALL PYINIT('CMS','e+','e-',500D0)    ! 500 GeV ILC
      CALL PYINIT('CMS','p','pbar',1960D0)   ! Tevatron
C      CALL PYINIT('CMS','p','p',14000D0)    ! 14 TeV LHC
C      CALL PYINIT('CMS','p','p',7000D0)     !  7 TeV LHC

C initialize histograms
      CALL PYBOOK(1,'ttbar invariant  mass',   100,0D0,1000D0)
      CALL PYBOOK(2,'PT distribution of top',  100,0D0,1000D0)
      CALL PYBOOK(3,'inv mass of b and e-',     50,0D0, 300D0)
      CALL PYBOOK(4,'inv mass of bbar and e-',  50,0D0, 300D0)
      CALL PYBOOK(5,'minimum sqrt(s)',         100,0D0,1000D0)
      
C-------------------------------------------
C...Second section: event loop.
C-------------------------------------------

C...Loop over the number of events.

      DO IEV=1,nevpythia
        IF(MOD(IEV,100).EQ.0) print*, 'Now at event number',IEV

C...Event generation.
        CALL PYEVNT
        CALL PYHEPC(1)  ! convert to HEPEVT standard
                        ! see section 5.4 of the manual

C...List first few events, say 2.
        IF(IEV.LE.2) then 
           CALL PYLIST(1)
           read*
        ENDIF

c...Example histogram: look at the ttbar invariant mass distribution
        do ihep=1,nhep   ! loop through all particles in the event record
           if(isthep(ihep).eq.3) then  ! only look at the summary portion
               if( idhep(ihep) .eq.  6 ) it    = ihep  ! found top 
               if( idhep(ihep) .eq. -6 ) itbar = ihep  ! found anti-top
               if( idhep(ihep) .eq. -11) ie    = ihep  ! found e-
               if( idhep(ihep) .eq. +11) iebar = ihep  ! found e+
               if( idhep(ihep) .eq.  5 ) ib    = ihep  ! found b
               if( idhep(ihep) .eq. -5 ) ibbar = ihep  ! found bbar
               if( idhep(ihep) .eq.  12) in    = ihep  ! found nu_e
               if( idhep(ihep) .eq. -12) inbar = ihep  ! found nu_ebar
           endif
        enddo      ! loop through all particles in the event record


C Fill histograms and calculate things.
        CALL PYFILL(1, inv_mass(it,itbar), 1D0)
        CALL PYFILL(2, pt(it),             1D0)
        CALL PYFILL(3, inv_mass(ie,ib),    1D0)    ! wrong pair
        CALL PYFILL(4, inv_mass(ie,ibbar), 1D0)    ! correct pair

cc  Smin
        do i=1,4
           pvis(i) = phep(i,ie)+phep(i,iebar)+phep(i,ib)+phep(i,ibbar)
        enddo     ! the total momentum of all visible particles

        ptvis = dsqrt( pvis(1)**2d0 + pvis(2)**2d0 )
                    ! PT sum of all visible particles
        mvis = pvis(4)**2d0 - pvis(1)**2d0 - pvis(2)**2d0 - pvis(3)**2d0
                    !  visible mass
        if(mvis .lt. 0d0) mvis=0d0
        mvis = dsqrt(mvis)

        smin = dsqrt( ptvis**2d0 + mvis**2d0 ) + ptvis

        CALL PYFILL(5, smin, 1D0)  
        
c Event selection of a CDF analysis for A_FB in dilepton channel
c
c CDF Note 10436, 5.1 fb-1
c A_fb = 0.42 ± 0.15stat ± 0.05syst
c A_fb(theory) = 0.06 ± 0.01
c
c electrons: CAL ET > 20 GeV (little energy in HCAL), 
c            |eta| < 1.1 and 1.2 < |eta| < 2.8
c muons: track PT > 20 GeV, |eta| < 1.0
c MET > 25 GeV
c jets: PT > 25 GeV, |eta| < 2.5
c HT > 200 GeV (HT scalar sum of MET, leptons, jets)

        MPT(1) = phep(1,in)+phep(1,inbar)
        MPT(2) = phep(2,in)+phep(2,inbar)
        MET = dsqrt( MPT(1)**2d0 + MPT(2)**2d0 )

        HT = MET + pt(ie) + pt(iebar) + pt(ib) + pt(ibbar) ! ignoring ISR

        if(       abs(eta(ie))   .lt.2.8 .and. pt(ie)   .gt.20d0  ! e-
     &      .and. abs(eta(iebar)).lt.2.8 .and. pt(iebar).gt.20d0  ! e+
     &      .and. abs(eta(ib))   .lt.2.5 .and. pt(ib)   .gt.25d0  ! b
     &      .and. abs(eta(ibbar)).lt.2.5 .and. pt(ibbar).gt.25d0  ! bbar
     &      .and.  MET           .gt.25d0                         ! MET
     &      .and.  HT            .gt.200d0                        ! HT
     & ) then
           counter = counter + 1

      endif
            
      ENDDO  ! Loop over the number of events.

C-------------------------------------------
C...Third section: produce output and end.
C-------------------------------------------

C...Cross section table.
      CALL PYSTAT(1)
      
C... Finalize analysis and report results
C... Plot histograms.
      
c PYDUMP(MDUMP,LFN,NHI,IHI)
c MDUMP=3: (x,y) format
c LFN: file number
c NHI: number of histograms to be dumped; 
c      if 0 then all existing histograms are dumped.
c IHI: array containing histogram numbers 
c      in the first NHI positions for NHI nonzero
      CALL PYDUMP(3,1,1,1)      ! ttbar inv mass
      CALL PYDUMP(3,2,1,2)      ! PT of the top
      CALL PYDUMP(3,3,1,3)      ! (e-,b)    - wrong pair
      CALL PYDUMP(3,4,1,4)      ! (e-,bbar) - correct pair
      CALL PYDUMP(3,7,1,5)      ! smin      
      
C print efficiency 

      print*, 'Total number of events before cuts: ', nevpythia
      print*, 'Total number of events after cuts : ', counter
      print*, 'efficiency                        : ', 
     &                       dble(counter)/dble(nevpythia)
      
\end{verbatim}
} %

The main code is divided into 3 sections; initialization, event loop and ending. 
In the first part, users set up process, the number of events, collider type, center of mass energy, 
beam environments, masses, decay patterns, initializing histograms etc.
The second part is for the event generation, where most of actual analysis is done.
Users are supposed to store relevant information of each event, since \pythia does not save event information by default.
Users fill histograms in this second part. We will use \pythia commands for plots but 
often many users use their own favorite plotting program and do not use plotting package in \pythia.
In the last section, histograms may be plotted, users can finalize analysis and report results. 
We will go over a few important commands below.

\begin{enumerate}

\item First of all, we save the number of events to generate in a variable, nevpythia.
Then we use MSEL variable to choose a process to be simulated. 
Selected other processes are shown below.
{\small
\begin{verbatim}
C...Standard Model
c      MSUB(1)=1    ! Drell-Yan
c      MSTP(43)=2   ! [1] only photon diagram included
c                   ! [2] only Z diagram included
c                   ! [3] both diagrams + interference
c      MSUB(18)=1   ! f fbar -> gamma gamma
c      CKIN(1)=100d0
c      MSUB(22)=1   ! f fbar -> Z Z
c      MSUB(23)=1   ! f fbar -> W Z
c      MSUB(24)=1   ! f fbar -> Z h
c      MSUB(25)=1   ! f fbar -> W+ W-
c      MSUB(26)=1   ! f fbar -> W h
c      MSUB(27)=1   ! f fbar -> h h

C...New physics: Doubly charged Higgs
c      MSUB(349)=1  ! f fbar -> HL++ HL--
c      MSUB(350)=1  ! f fbar -> HR++ HR--
c      MSUB(351)=1  ! f f    -> f f HL 
c      MSUB(352)=1  ! f f    -> f f HR 
c      kc = pycomp(9900041)
c      PMAS(kc,1)=1000
c      kc = pycomp(9900042)
c      PMAS(kc,1)=1000

c...top quark
      MSEL=6     ! t quark 

C...New physics: fourth generation/Little Higgs
c...t' 
c      MSEL=8         ! fourth generation t' pair production
c      MSTP(7)=8      ! choice of heavy flavor, superseeded by MSEL=4-8
c      PMAS(8,1)=500  ! mass of the t'
c      MSEL=38        ! fourth generation single t' 
c      MSUB(83)=1     ! q f -> Q f
c...b' 
c      MSEL=7         ! fourth generation b' pair production
c      MSTP(7)=7      ! choice of heavy flavor, superseeded by MSEL=4-8
c      PMAS(7,1)=400  ! mass of the b'
c      MSEL=37        ! fourth generation single b' 
c      MSUB(83)=1     ! q f -> Q f 

c      MSTP(1)=4      ! four generations
c      do i=56,75     ! Turn on b' and t' decays
c        mdme(i,1) = 1   
c      enddo
c      MSTP(127)=1    ! do not crash if vanishing cross-sections
\end{verbatim}
}
\item Users are allowed to force a particular decay of certain particle, $W$ in this example.
This is done with ``mdme'' switch. To list all decay channels, widths etc, one can use ``CALL PYSTAT(2)''.

\item PMAS sets masses.
\item Initial and final state radiation, multiple interactions and fragmentation/hadronization are treated by 
MSTP switches.

\item To initialize the collider type, use ``PYINIT'', for instance, ``CALL PYINIT(CMS,p,pbar,1960D0)'' for Tevatron.

\item Event generation is performed by calling ``PYEVNT'' and information can saved into HEPEVT standard format for further analysis.

\item ``CALL PYLIST(1)'' shows basic event listing, which is self-explanatory.
{\scriptsize
\begin{verbatim}
                            Event listing (summary)

I particle/jet KS     KF  orig    p_x      p_y      p_z       E        m

 1 !p+!        21   2212    0    0.000    0.000  980.000  980.000    0.938 
 2 !pbar-!     21  -2212    0    0.000    0.000 -980.000  980.000    0.938
==============================================================================
 3 !u!         21      2    1    0.000    0.000  479.080  479.080    0.000
 4 !ubar!      21     -2    2    0.000    0.000 -137.794  137.794    0.000
 5 !u!         21      2    3    0.000    0.000  479.080  479.080    0.000
 6 !ubar!      21     -2    4    0.000    0.000 -137.794  137.794    0.000
 7 !t!         21      6    0  -39.462  145.528   32.386  232.316  173.741
 8 !tbar!      21     -6    0   39.462 -145.528  308.899  384.558  172.426
 9 !W+!        21     24    7  -65.111  123.272   76.449  176.479   76.583
10 !b!         21      5    7   25.648   22.256  -44.063   55.837    4.800
11 !W-!        21    -24    8   77.947  -79.254  123.668  184.042   78.871
12 !bbar!      2      -5    8  -38.485  -66.274  185.231  200.517    4.800
13 !e+!        21    -11    9   -3.487   85.329   30.144   90.564    0.001
14 !nu_e!      21     12    9  -61.624   37.943   46.305   85.915    0.000
15 !e-!        21     11   11   69.103  -30.988  109.185  132.879    0.001
16 !nu_ebar!   21    -12   11    8.844  -48.266   14.483   51.163    0.000
==============================================================================
\end{verbatim}
}

\item It is highly recommend to save particle IDs in some variables as shown in the example.
This is done using ``idhep'' command and PDG numbers.

\item Users can compute certain quantities and do analysis afterwards. In this example, we are making 5 histograms, 
invariant mass of the top pair, transverse momentum of the top, invariant mass of a lepton and $b$/$\bar{b}$, and $\sqrt{s}_{min}$.

\item In the last section of the main code, users can produce output/figures and end the program. 
For instance, the output of this example is shown below. The cross section table is called by using ``CALL PYSTAT(1)''. 
The second part is efficiency due to some cuts that are introduced. It shows that 51\% of total 10K events passed those cuts.
{\scriptsize
\begin{verbatim}
********* PYSTAT:  Statistics on Number of Events and Cross-sections *********

 ==============================================================================
 I                                  I                            I            I
 I            Subprocess            I      Number of points      I    Sigma   I
 I                                  I                            I            I
 I----------------------------------I----------------------------I    (mb)    I
 I                                  I                            I            I
 I N:o Type                         I    Generated         Tried I            I
 I                                  I                            I            I
 ==============================================================================
 I                                  I                            I            I
 I   0 All included subprocesses    I        10000        201243 I  7.174D-11 I
 I  81 q + qbar -> Q + Qbar, mass   I         9402        185523 I  6.757D-11 I
 I  82 g + g -> Q + Qbar, massive   I          598         15720 I  4.170D-12 I
 I                                  I                            I            I
 ==============================================================================

 ********* Total number of errors, excluding junctions =        0 *************
 ********* Total number of errors, including junctions =        0 *************
 ********* Total number of warnings =                           0 *************
 ********* Fraction of events that fail fragmentation cuts =  0.00000 *********

 Total number of events before cuts:        10000
 Total number of events after cuts :         5114
 efficiency                        :   0.51139999999999997     
\end{verbatim}
}
The other outputs are figures. This example uses \pythia commands (``PYDUMP''), to write results into files.
\end{enumerate}

\begin{figure}[t]
\centerline{    \epsfig{file=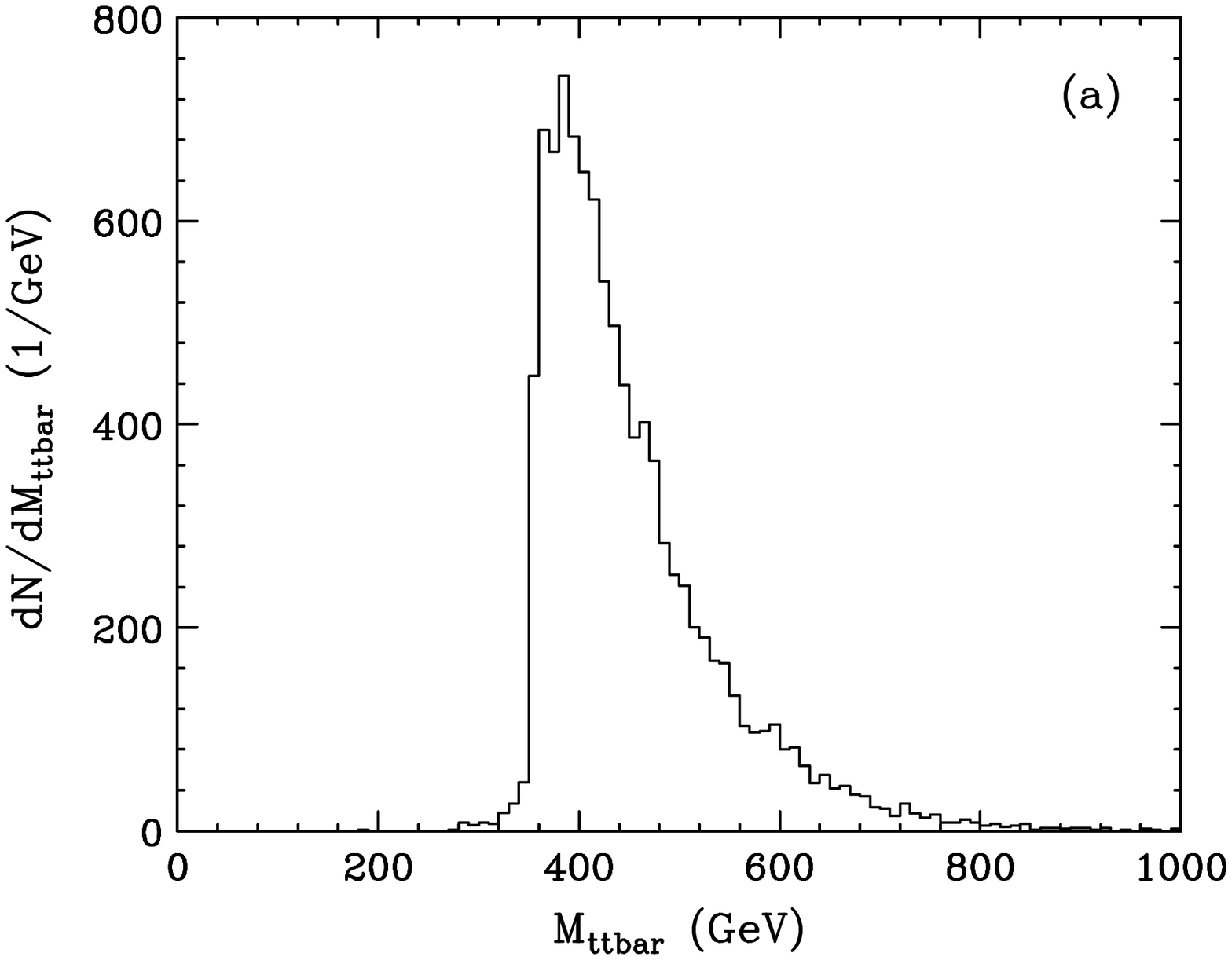,width=2.3in}\hspace{0.1cm} \epsfig{file=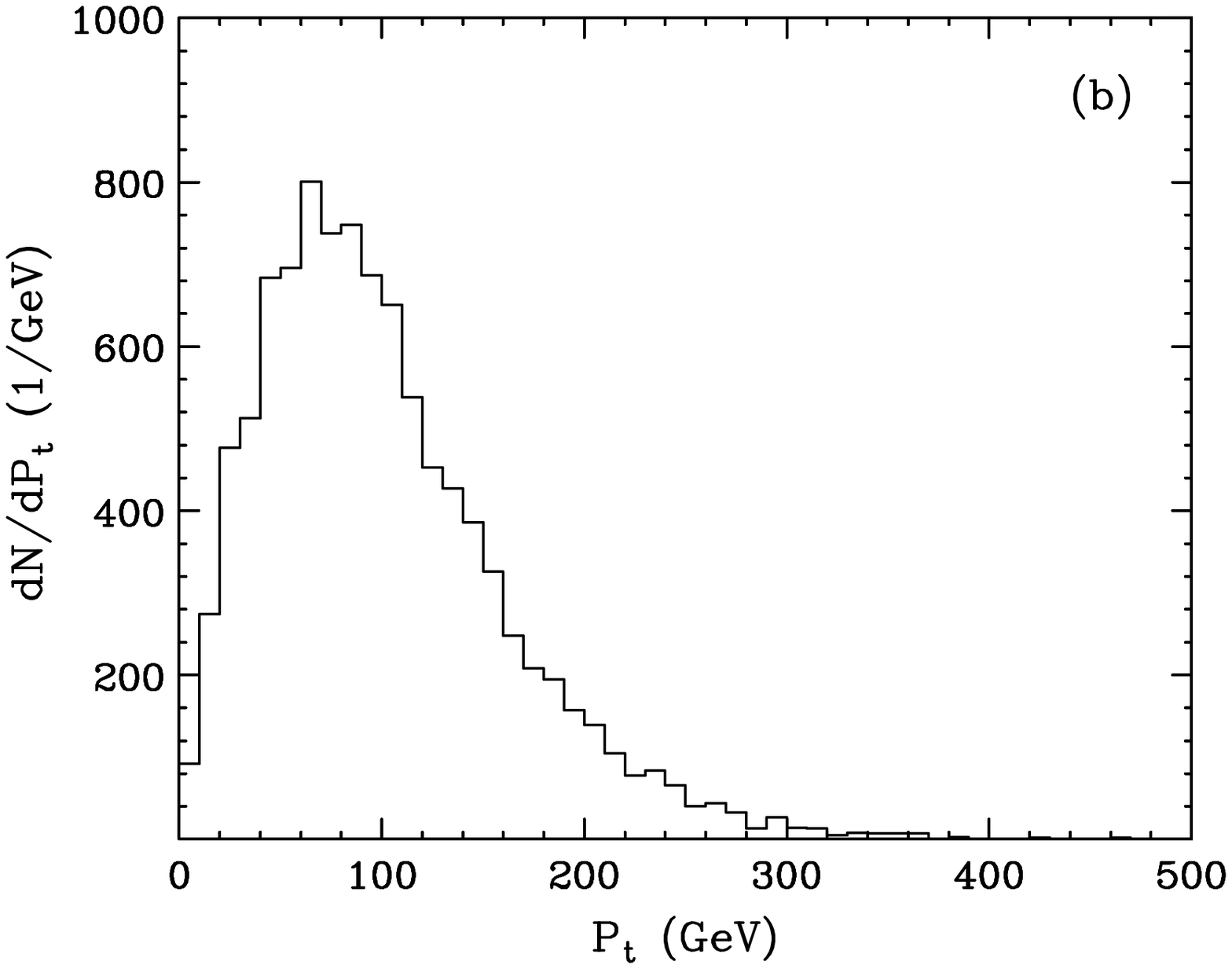,width=2.3in} } 
\vspace{0.4cm}
\centerline{    \epsfig{file=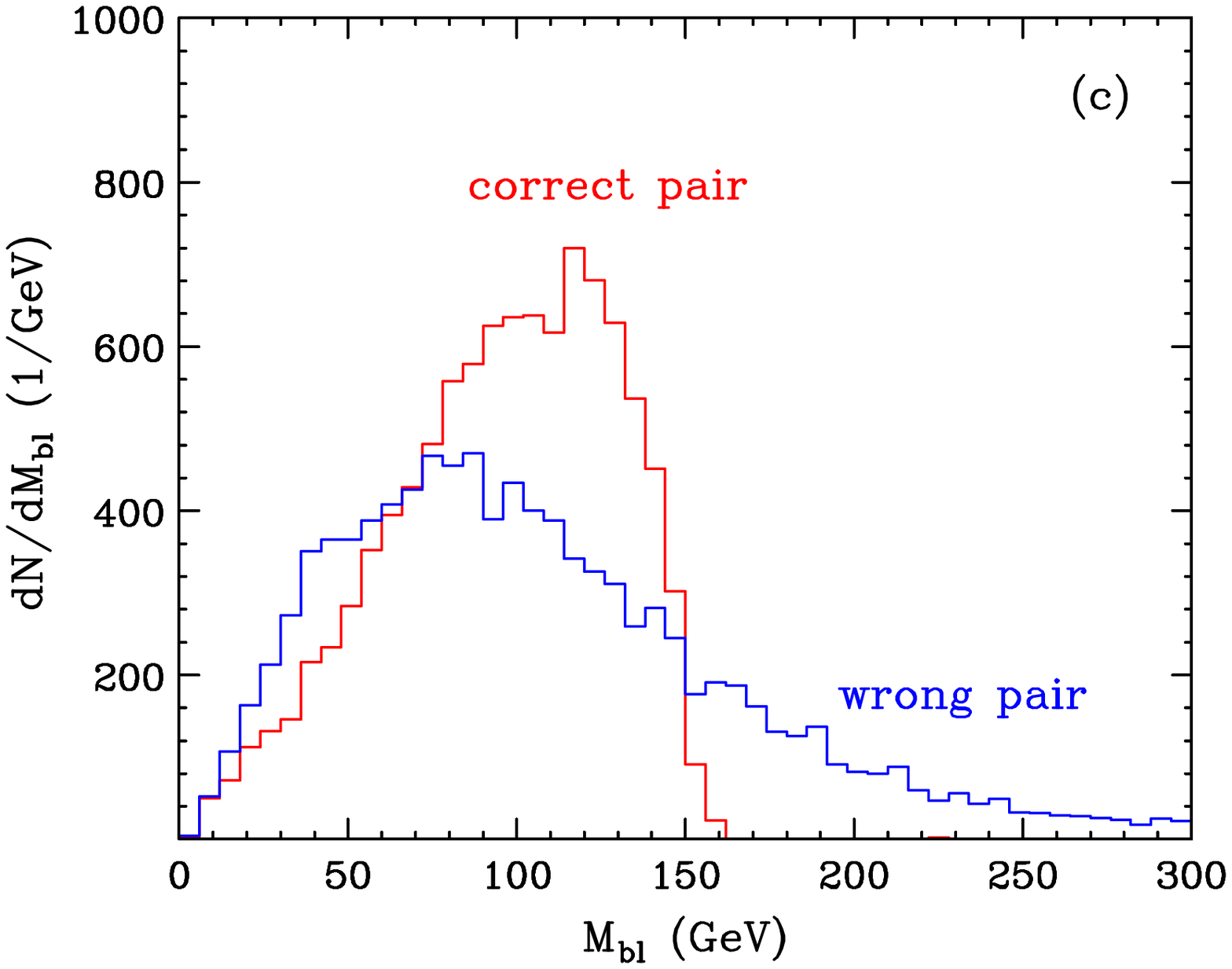,width=2.3in} \hspace{0.1cm} \epsfig{file=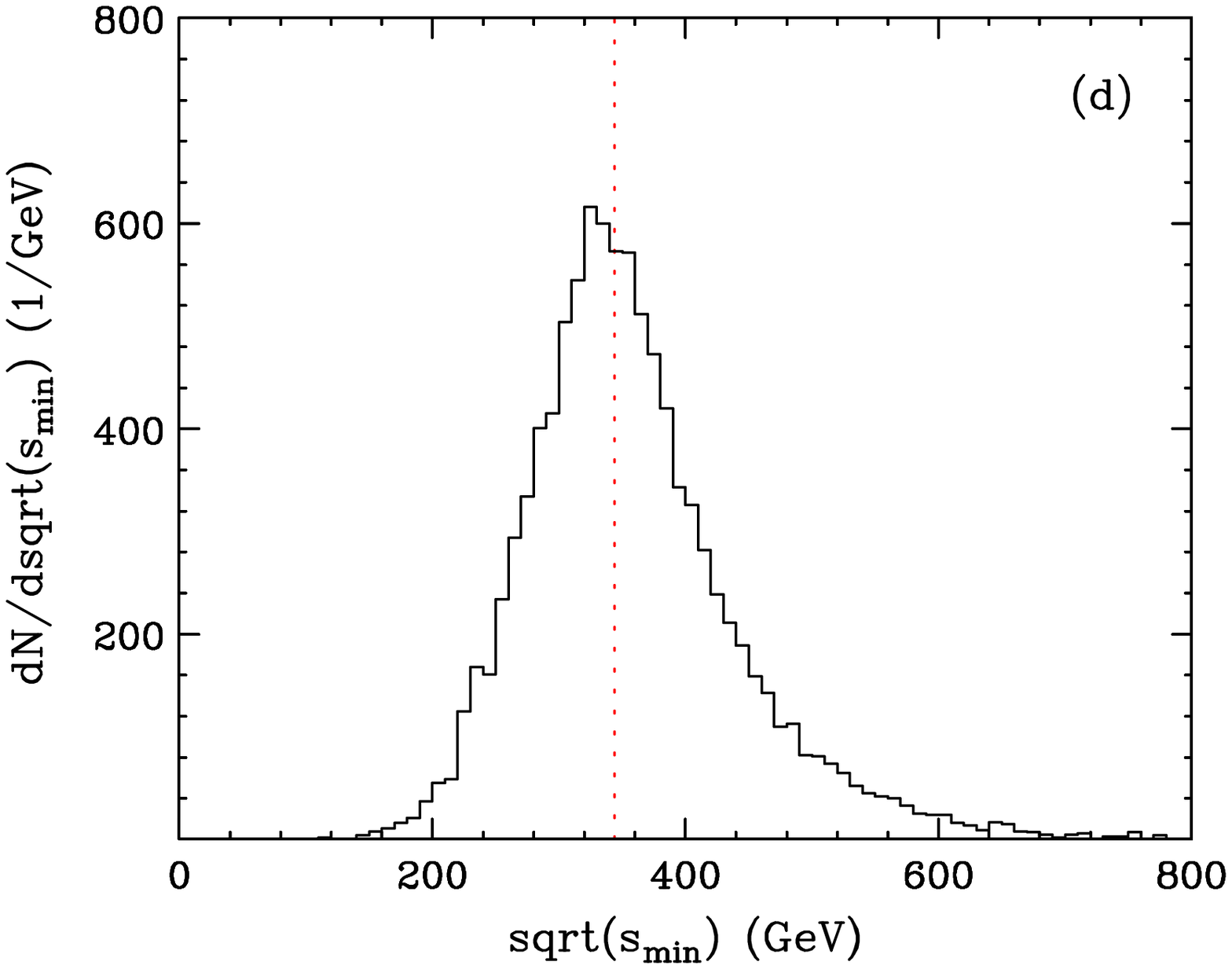,width=2.3in} }
\caption{Various distributions in the dilepton channel of the $t\bar{t}$ production at the Tevatron. (a) $M_{t\bar{t}}$ (b) $P_t$ 
(c) $M_{b\ell}$ (d) $\sqrt{s}_{min}$.}
\label{fig:ttbar}
\end{figure}

With help of a plotting program, we show the following results. 
Fig. \ref{fig:ttbar} shows various distributions in the $t\bar{t}$ dilepton production:
$t\bar{t}$ invariant mass (a), transverse momentum of the top (b), 
invariant mass of a b-quark and a lepton (c), and $\sqrt{s}_{min}$ (d). 
In Fig. \ref{fig:ttbar}(c), both wrong and correct combinations are shown, which are different and therefore 
can be used to reduce combinatorial background \cite{Baringer:2011nh}. 
Even in the case of two missing neutrinos the $\sqrt{s}_{min}$ exhibits a peak at $2 M_t$, which is shown in the vertical line \cite{Konar:2008ei,Konar:2010ma}. 
%

%%%%%%%%%%%%%%%%%%%%%%%%%%%%%%%%%%%%%%%%%%%%%%%%%%
\subsubsection{slepton production}
%%%%%%%%%%%%%%%%%%%%%%%%%%%%%%%%%%%%%%%%%%%%%%%%%%

\begin{figure}[t]
\centerline{    \epsfig{file=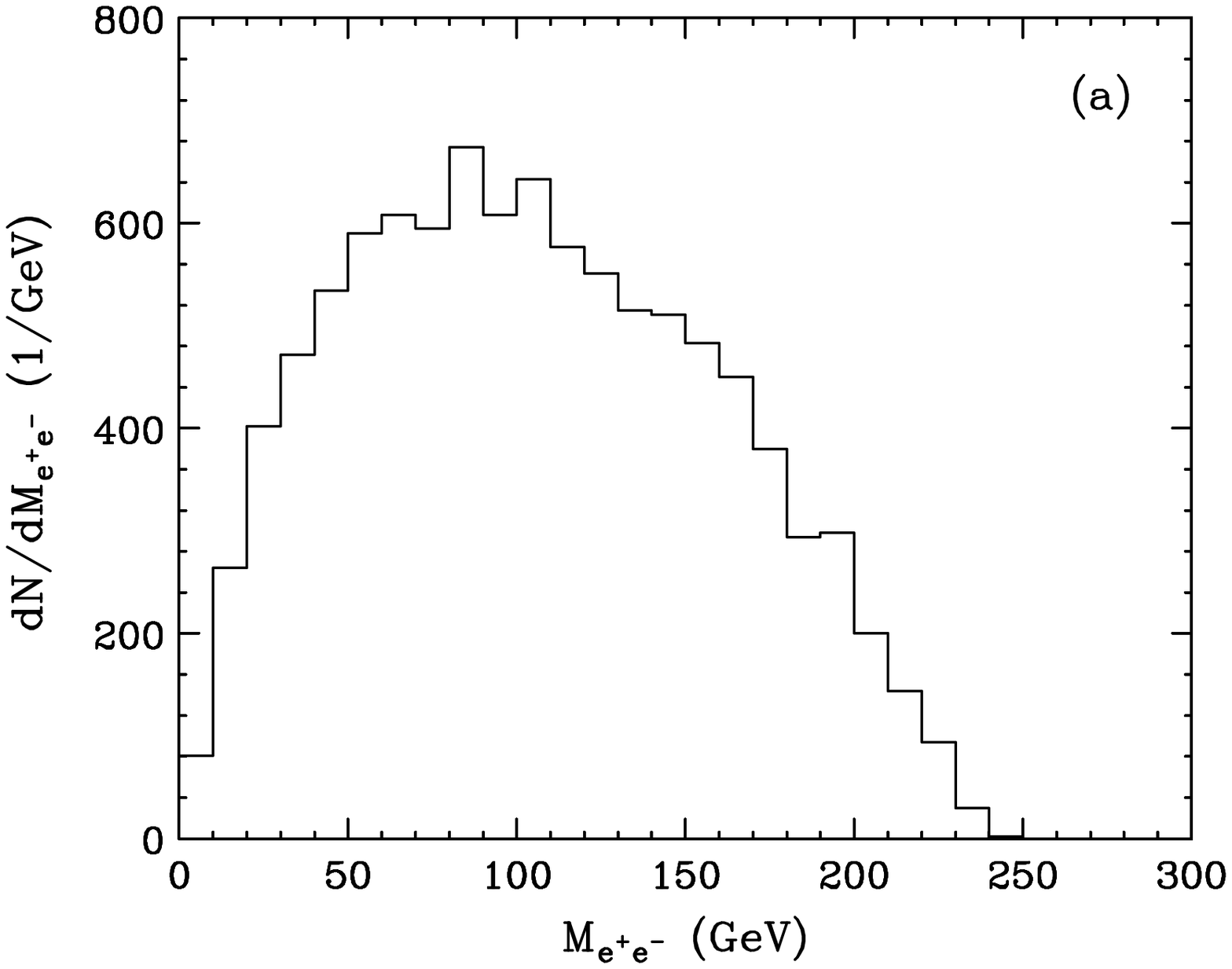,width=2.3in} \epsfig{file=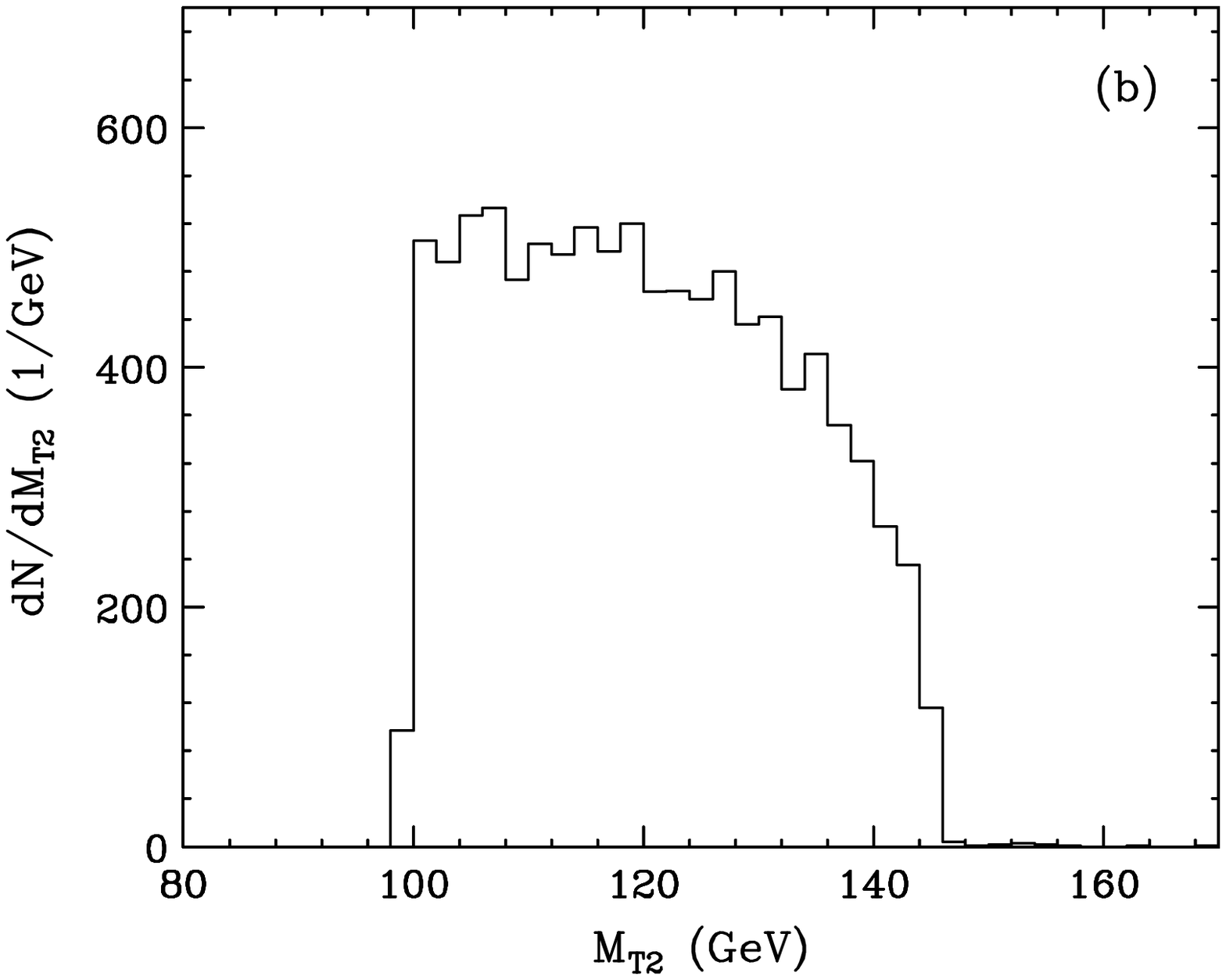,width=2.3in} } 
\vspace{0.3cm}
\centerline{    \epsfig{file=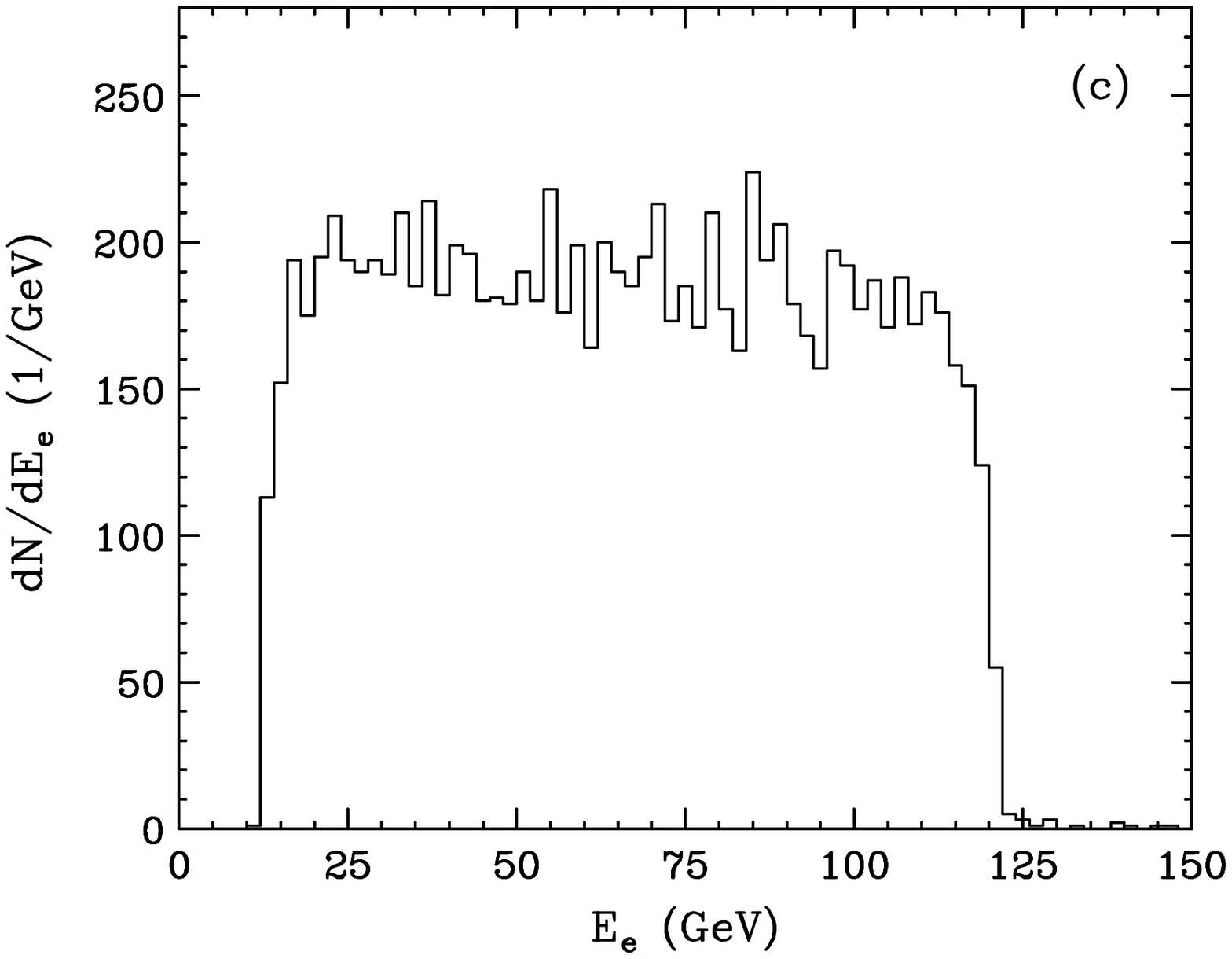,width=2.3in}        \epsfig{file=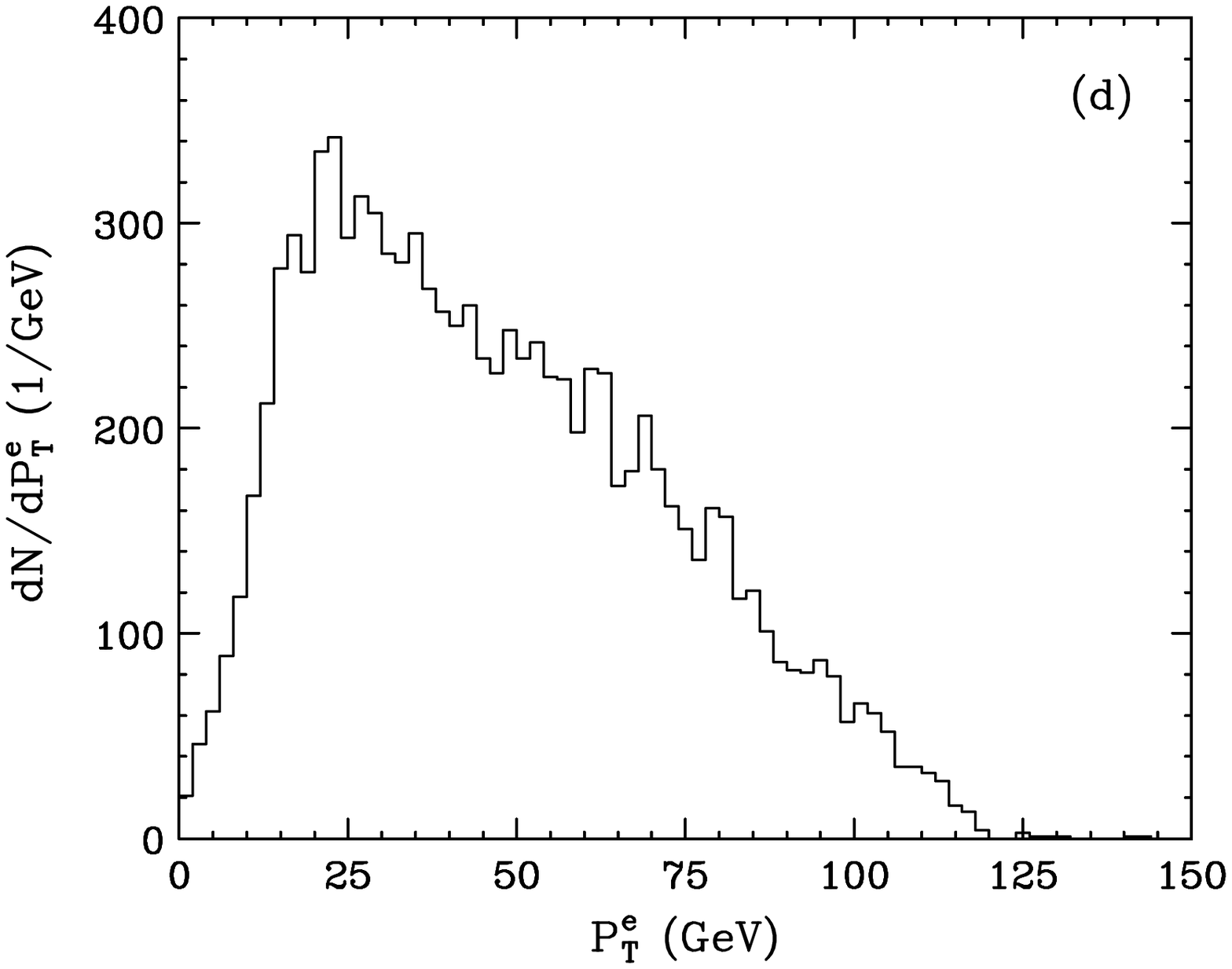,width=2.3in}         }
\caption{
Various distributions for selectron production at a 500 GeV ILC. (a) $M_{e^+ e^-}$ (b) $M_{T2}$ 
(c) $E_e$ (d) $P_T^e$.
}
\label{fig:slepton}
\end{figure}

In this example, we consider pair production of the right handed selectron at an 500 GeV ILC.
\begin{enumerate}
\item Main structure of the code is very similar to the previous example. 
The process is again set up by ``MSEL''. In this case we use ``MSUB(202)=1''. 
Other examples of SUSY processes are shown below and a complete list is found in the manual. 
Users have various options to define mass spectrum for SUSY production: \pythia RG running, external mass spectrum generators or LHA input files.
We will consider one of mSugra point, SPS1a, and use \pythia commands, ``IMSS'' and ``RMSS'', which define mass spectrum.
\begin{verbatim}
C...Generic SUSY simulation (SUGRA scenario)
c      MSEL=39       ! All MSSM processes at once
c      MSEL=42       ! Slepton production
c      MSEL=0        ! One by one
      MSUB(202)=1   ! f fbar -> ~e_R  ~e_Rbar
c      MSUB(201)=1   ! f fbar -> ~e\_L  ~e\_Lbar
c      MSUB(204)=1   ! f fbar -> ~mu\_L ~mu\_Lbar
c      MSUB(208)=1   ! f fbar -> ~tau2 ~tau2bar
c
c      MSUB(210)=1   ! f fbar -> ~l\_L ~nu\_Lbar
c      MSUB(212)=1   ! f fbar -> ~tau\_2 ~nutaubar
c
c      MSUB(213)=1   ! f fbar -> ~nul ~nulbar
c      MSUB(214)=1   ! f fbar -> ~nutau ~nutaubar
c
      IMSS(1)=2     ! MSUGRA spectra from analytic approximation
c      IMSS(1)=12    ! MSUGRA spectra from IsaJet
      RMSS(1)=250D0 ! Mhalf: common gaugino mass
      RMSS(4)=1D0   ! Sign of mu (magnitude irrelevant if IMSS(1)=2)
      RMSS(5)=10D0  ! tan beta
      RMSS(8)=100D0 ! M0: common scalar mass
      RMSS(16)=0D0  ! A0: common trilinear coupling
\end{verbatim}
\item The collider type and energy are set by ``PYINIT''.
\begin{verbatim}
      CALL PYINIT('CMS','e+','e-',500D0)    ! 500 GeV ILC
\end{verbatim}
\item The rest is basically the same as all other examples. 
Initializing histograms, identifying particles, calculating relevant quantities, 
filling histograms, calculating efficiency, plotting histograms, report results and etc.
For instance, to calculate $M_{T2}$, one needs momentum of visible particles and the missing transverse momentum.
In this case, two visible particles are electron and positron, and the missing momentum is balanced by the momentum of $e^+e^-$ pair.
\begin{verbatim}
  do ihep=1,nhep  ! loop through all particles in the event record
     if(isthep(ihep).eq.3) then ! only look at the summary portion            
           if( idhep(ihep) .eq. -11) ie    = ihep ! found e-
           if( idhep(ihep) .eq.  11) iebar = ihep ! found e+
     endif
  enddo           ! loop through all particles in the event record

  pin1(1) = phep(1,ie)
  pin1(2) = phep(2,ie)
  pin2(1) = phep(1,iebar)
  pin2(2) = phep(2,iebar)
  mchi    = PMAS(310,1)  ! KC =  PYCOMP(1000022)
  mt2 = mt2_sleptons(pin1,pin2,mchi)
\end{verbatim}
The four momenta of particles are saved in ``PHEP'' variable. 
\end{enumerate}
Various kinematic distributions are shown in Fig. \ref{fig:slepton}:
the invariant mass of the $e^+e^-$ (a), 
$M_{T2}$ (b), 
the energy distribution (c) and the transverse momentum (d) of the electron.
The energy distribution and the $M_{T2}$ exhibit interesting end point structures which are useful for 
extracting particles masses, sleptons and neutralinos in this example \cite{Barr:2010zj,Barr:2011xt}. 
%

%%%%%%%%%%%%%%%%%%%%%%%%%%%%%%%%%%%%%%%%%%%%%%%%%%
\subsubsection{$\rho_T \to \pi_T + W$}
%%%%%%%%%%%%%%%%%%%%%%%%%%%%%%%%%%%%%%%%%%%%%%%%%%

\begin{figure}[t]
\centerline{    \epsfig{file=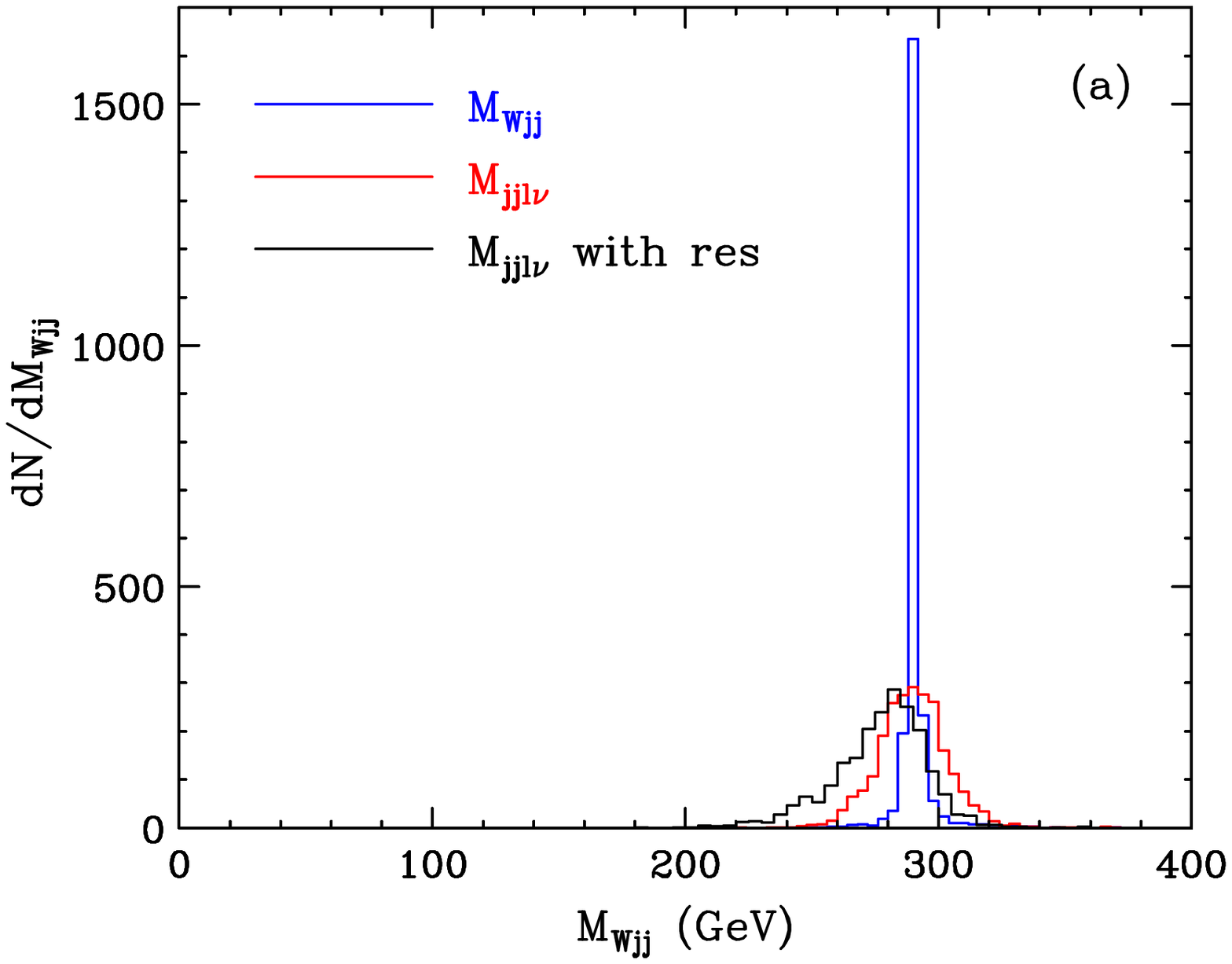,width=2.3in} 
                        \epsfig{file=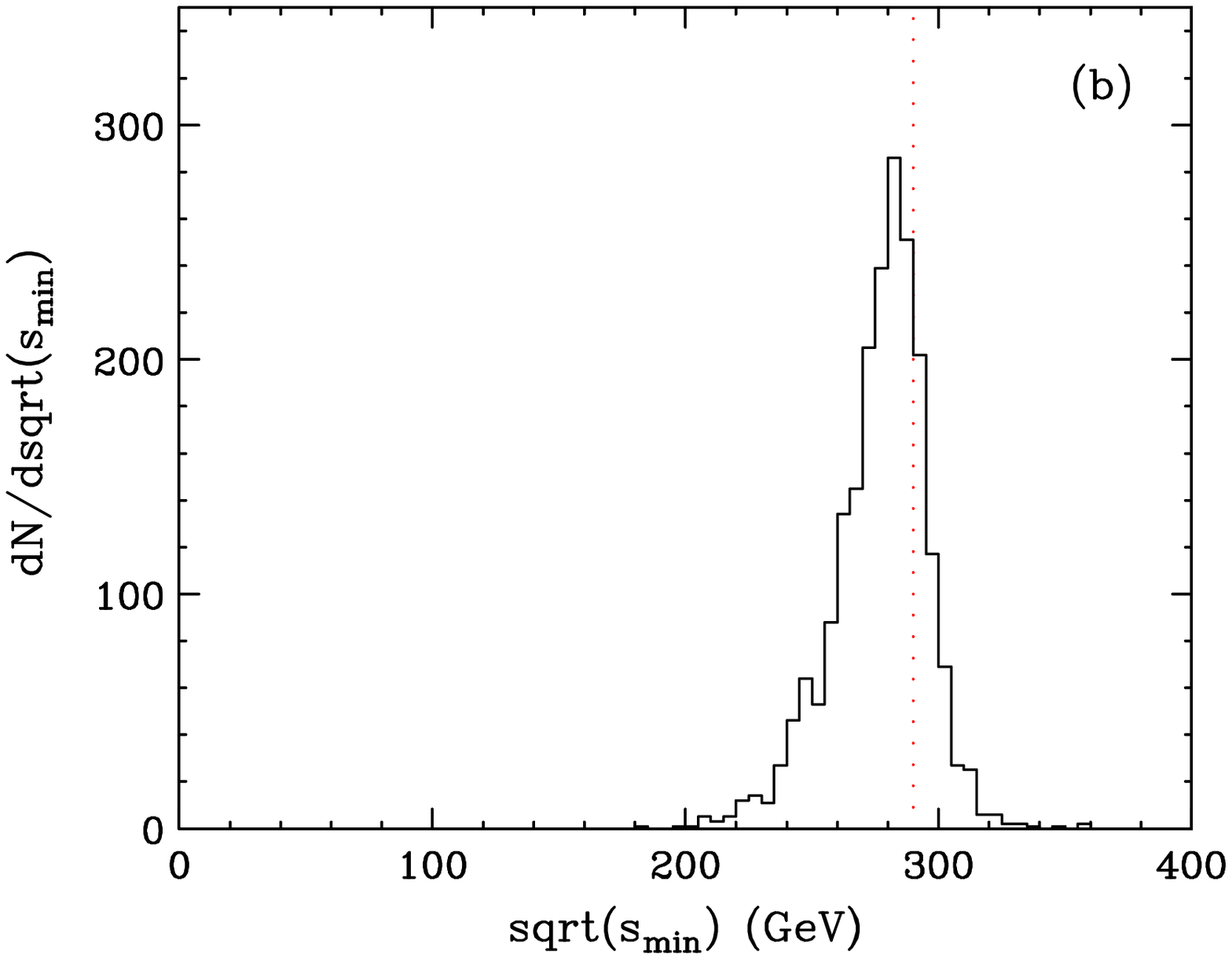,width=2.3in}            }
\caption{Invariant mass distribution (a) and $\sqrt{s}_{min}$ (b) for $\rho_T^0 \to \pi_T^\pm + W^\mp \to jj + \ell^\mp \nu_\ell$ at the Tevatron.}
\label{fig:techni}
\end{figure}

In this example, we simulate techni $\rho$ production with {\tt PYTHIA} at the Tevatron, setting masses 
$M_{\rho_T^0} = 290$ GeV and $M_{ \pi_T^\pm}$ = 160 GeV. 
We consider the $\rho_T^0 $ decay to $\pi_T^\pm + W^\mp$ and $\pi_T^\pm$ to 2 jets. 
For faster event generation, we force the decay of $W$ into electron and neutrino. 
\begin{verbatim}
C...Possibility to set masses freely:
C...pi_tech0
      PMAS(PYCOMP(KTECHN+111),1)=160D0   !   KTECHN=3000000
C...pi_tech+-.                           !   for techi-particles
      PMAS(PYCOMP(KTECHN+211),1)=160D0
C...rho_tech0
      PMAS(PYCOMP(KTECHN+113),1)=290D0

C...rho_tech0.
      MSUB(191)=1
C...rho_tech+-. 
c      MSUB(192)=1
C...omega_tech.
c      MSUB(193)=1

C...Sample code to force only decays of interest
      do i=190,208    
        mdme(i,1) = 0   ! Turn off all W decays
      enddo
      mdme(206,1) = 1   ! Turn on electron + neutrino
\end{verbatim}
Forming an invariant mass of the two jets, one can easily find a bump. 
With fully reconstructed $W$, the  $\rho_T^0 $ appears as a resonance in the invariant mass of $W+jj$, 
which is shown as the blue histogram in Fig. \ref{fig:techni}(a).
In the leptonic decay of the $W$, the transverse momentum of the neutrino is determined by the 
missing transverse momentum, assuming that the neutrino is the only missing particle in the event.
Mass-shell condition of the $W$ provides the $z$-component of the neutrino momentum up to two fold ambiguity. 
Invariant mass of two jets and $\ell \nu_\ell$ system with both solutions is shown in red.
With appropriate detector effects, the invariant mass get further smeared as shown in back in Fig. \ref{fig:techni}(a).
Without reconstructing the neutrino momentum, one can obtain mass information 
looking at the end point structure in the $\sqrt{s}_{min}$ distribution as shown in Fig. \ref{fig:techni}(b).
The red vertical lines shows location of $\rho_T^0$ resonance.
%

%%%%%%%%%%%%%%%%%%%%%%%%%%%%%%%%%%%%%%%%%%%%%%%%%%
\subsubsection{The same sign dilepton in SUSY: LM6}
%%%%%%%%%%%%%%%%%%%%%%%%%%%%%%%%%%%%%%%%%%%%%%%%%%

\begin{figure}[t]
\centerline{    \epsfig{file=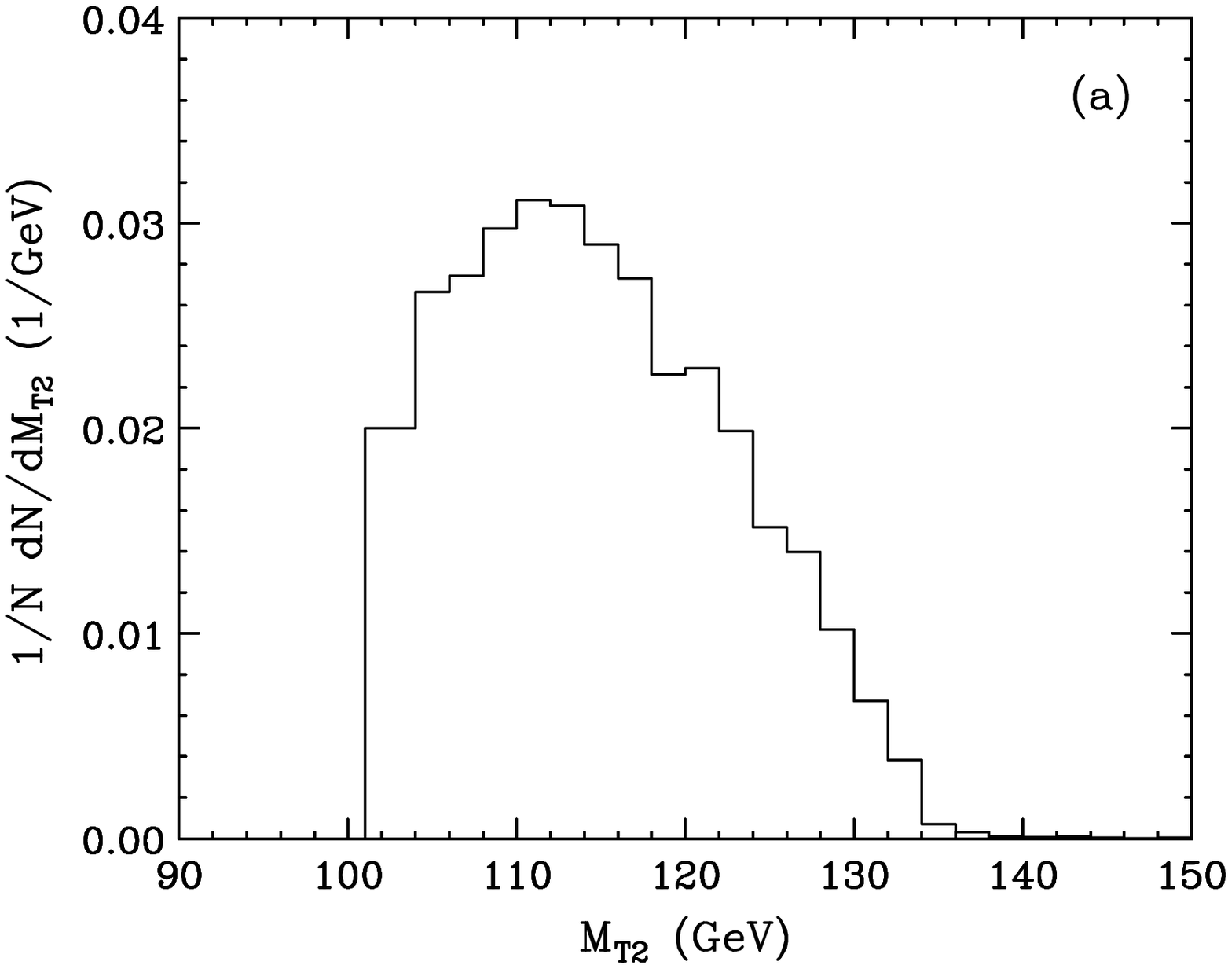,width=2.3in} \epsfig{file=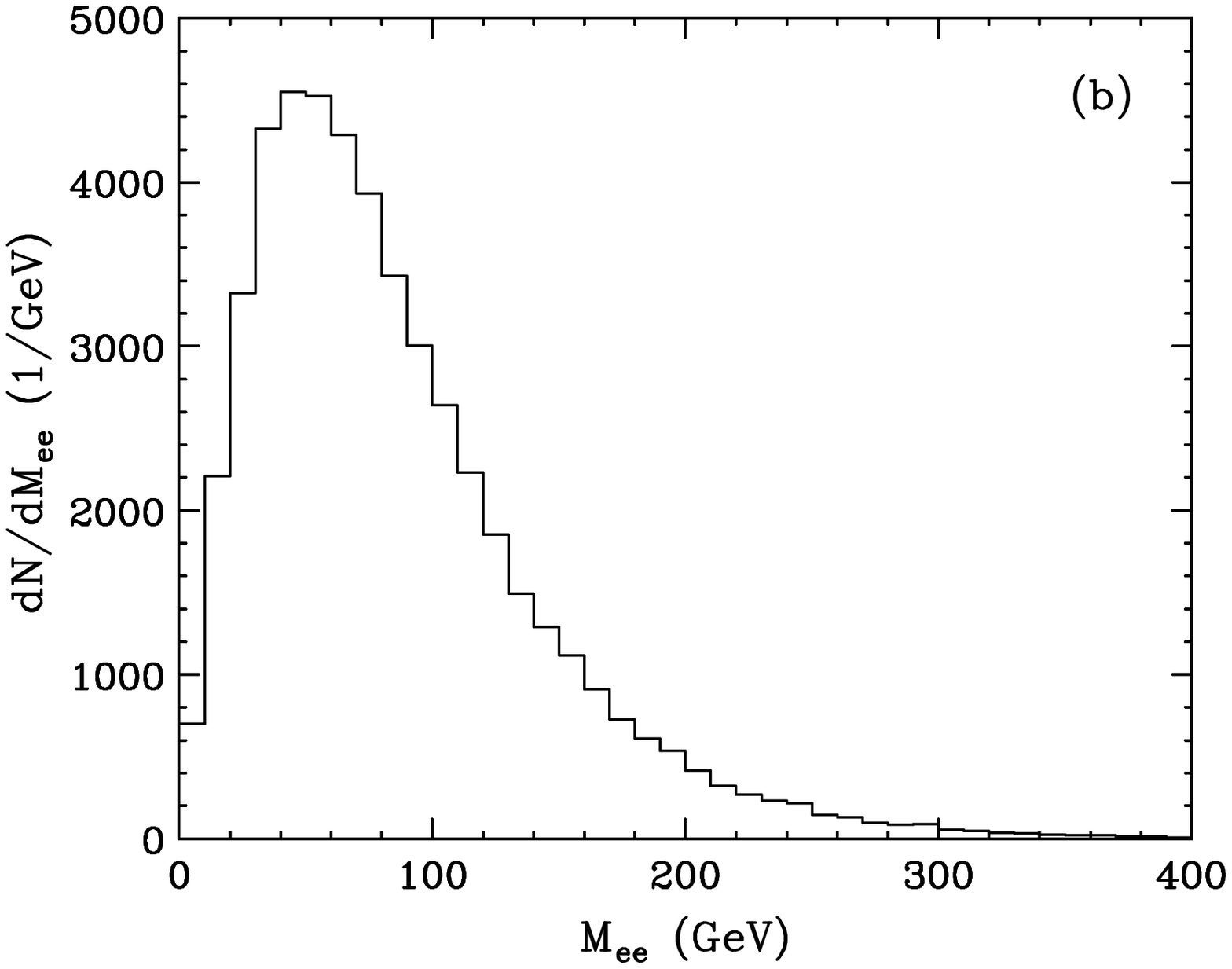,width=2.3in} }      
\caption{$M_{T2}$ (a) and invariant mass (b) distributions of the same-sign dilepton for LM6 at the 14 TeV LHC.}
\label{fig:LM6}
\end{figure}

\begin{figure}[t]
\centerline{    \epsfig{file=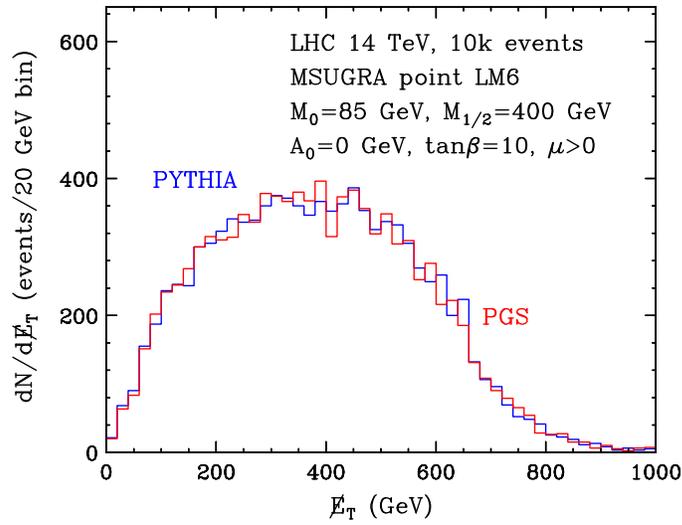,width=3.5in}            }
\caption{Missing transverse momentum distribution $\met$ for LM6.}
\label{fig:LM6PGS}
\end{figure}

We consider a pair production of gluino for LM6 point (a mSugra point in CMS) at the 14 TeV LHC. 
This time we define mass spectrum from a file in LHA format, which is generated by SuSpect\cite{Djouadi:2002ze}.
This is done with ``IMSS'' and ``PYSLHA'', as shown below.
{\small
\begin{verbatim}
      MSUB(244) = 1      ! g g    ->   gluino gluino

c-------------------------------------------------------------------
c      IMSS(1)=2     ! MSUGRA spectra from analytic approximation
cc      IMSS(1)=12    ! MSUGRA spectra from IsaJet
c      RMSS(1) =400D0 ! Mhalf: common gaugino mass
c      RMSS(4) =1D0   ! Sign of mu (magnitude irrelevant if IMSS(1)=2)
c      RMSS(5) =10D0  ! tan beta
c      RMSS(8) =85D0 ! M0: common scalar mass
c      RMSS(16)=0D0  ! A0: common trilinear coupling

      open(99,FILE='suspect.txt')
      IMSS(1) = 11    ! for SLHA input 
      IMSS(13) = 0    ! 0=MSSM particle content, 1=NMSSM
      IMSS(21) = 99   ! Logical Unit Number for SLHA spectrum read-in.
      IMSS(22) = 99   ! Logical Unit Number for SLHA decay read-in.
                      ! normally this should go together for consistency
      CALL PYSLHA(5,0,IFAIL)
      CALL PYSLHA(2,0,IFAIL)
      close(99)
C...MUPDA=0 : READ QNUMBERS/PARTICLE ON LUN=IMSS(21)
C...MUPDA=1 : READ SLHA SPECTRUM ON LUN=IMSS(21)
C...MUPDA=2 : LOOK FOR DECAY TABLE FOR KF=KFORIG ON LUN=IMSS(22)
C...          (KFORIG=0 : read all decay tables)
C...MUPDA=3 : WRITE SPECTRUM ON LUN=IMSS(23)
C...MUPDA=4 : WRITE DECAY TABLE FOR KF=KFORIG ON LUN=IMSS(24)
C...MUPDA=5 : READ MASS FOR KF=KFORIG ONLY
C...          (KFORIG=0 : read all MASS entries)
 
c-----------------------------------------------------------------
c      IMSS(1) = 1  ! A general MSSM simulation
c      RMSS( 1) = 100.0         ! M_1 500.0'
c      RMSS( 2) = 300.0        ! M_2
c      RMSS( 3) = 1225.0       ! M_3 
c      RMSS( 4) = 1100.0        ! Mu 200.0'
c      RMSS( 5) =  10.0         ! Tan_beta
c      RMSS( 6) = 1800.0         ! M_Sl_L
c      RMSS( 7) = 1800.0         ! M_Sl_R
c      RMSS( 8) =5500.0        ! M_Sq_L
c      RMSS( 9) =5500.0        ! M_Sq_R
c      RMSS(10) =5500.0        ! M_Sq3_L
c      RMSS(11) =5500.0        ! M_Sbottom_R
c      RMSS(12) =5500.0        ! M_Stop_R
c      RMSS(13) =5500.0        ! M_Stau_L
c      RMSS(14) =5500.0        ! M_Stau_R
c      RMSS(15) = 800.0        ! A_b=Bottom trilinear coupling
c      RMSS(16) = 800.0        ! A_t=Top trilinear coupling
c      RMSS(17) =   0.0        ! A_tau=Tau trilinear coupling
c      RMSS(19) = 400.0        ! M_A=Psc higgs param.
c-----------------------------------------------------------------      
      
      CALL PYINIT('CMS','p','p',14000D0)    ! 14 TeV LHC
\end{verbatim}
}
Fig. \ref{fig:LM6} shows the 1D decomposed $M_{T2}$ (a) and invariant mass (b) distributions of the same-sign dilepton for LM6 at the 14 TeV LHC 
and Fig. \ref{fig:LM6PGS} shows the distribution of the missing transverse momentum with and without detector effects.
%

%%%%%%%%%%%%%%%%%%%%%%%%%%%%%%%%%%%%%%%%%%%%%%%%%%
\subsection{CalcHEP-PYTHIA Interface}
%%%%%%%%%%%%%%%%%%%%%%%%%%%%%%%%%%%%%%%%%%%%%%%%%%

\calchep is often linked to \pythia for further simulation, especially to include ISR, FSR, multiple parton interaction and etc.
To use this feature, we need an event-file in LHA format, which is generated by \calchep, and will be fed into \pythia.
Suppose the file name is ``event\_mixer.lhe'', which we have generated in an earlier example, using ``event\_mixer''.
We also need ``event2pyth.c'', which is in the \calchep ``utile/'' directory\cite{Belyaev:2000wn}. 
Here we should make sure a routine ``UPEVNT'' is deleted in this ``event2pyth.c'', since \pythia already has one.
The rest is to write a main code for \pythia, part of which is shown below.
{\small
\begin{verbatim}
C...If interested only in cross sections and resonance decays: 
C...switch off initial and final state radiation, 
C...multiple interactions and hadronization.
      MSTP(61)=0  ! initial state radiation [0] off, [D=2] on
                  ! [1]: on for QCD radiation in hadronic events and
                  !             QED radiation in leptonic ones
                  ! [2]: on for QCD/QED radiation in hadronic events and
                  !             QED radiation in leptonic ones
      MSTP(71)=0  ! final state radiation [0] off, [D=1] on
      MSTP(81)=0  ! multiple interactions [0] off, [D=1] on
      MSTP(111)=1 ! fragmentation and decay [0] off, [D=1] on
      MSTP(91)=0  ! No primordial kT
      
c...USER MODE
      MSTP(161)=21
      MSTP(162)=21

      OPEN(21, FILE='event_mixer.lhe',STATUS='UNKNOWN')

      CALL PYINIT('USER',' ',' ',0d0)
\end{verbatim}
}
Two Fortran files and one C file need to be compiled together. Make sure to include ``event2pyth.o'' in your makefile.
\begin{verbatim}
OBJS = example_interface.o ../pythia-6.4.25.o event2pyth.o
\end{verbatim}
%

%%%%%%%%%%%%%%%%%%%%%%%%%%%%%%%%%%%%%%%%%%%%%%%%%%
\section{Summary}
%%%%%%%%%%%%%%%%%%%%%%%%%%%%%%%%%%%%%%%%%%%%%%%%%%

In high energy physics there are various tools for different purposes.
None of these tools is perfect and 
users should know their advantages and disadvantages before choosing one for their study.
We have looked at \calchep and \pythia among many tools.
Both are commonly used event generators.
Especially \calchep is linked to \micromegas for dark matter study and \pythia is 
a hardcore software for collider physics. 
Although we only caught a glimpse of what they can do, 
it is our hope that beginners get basic ideas behind complicated structures and are not afraid of using them, and 
more advanced users get usefulness out of specific examples. 
\calchep and \pythia programs are continuously being improved and many current issues will be addressed in the future update.

%%%%%%%%%%%%%%%%%%%%%%%%%%%%%%%%%%%%%%%%%%%%%%%%%%
\section*{Acknowledgments}
%%%%%%%%%%%%%%%%%%%%%%%%%%%%%%%%%%%%%%%%%%%%%%%%%%
I would like to thank all the TASI students for their questions, and 
the organizers, Professors K. Matchev, T. Tait, T. DeGrand and K. T. Mahanthappa for their hospitality and great atmosphere. 
Special thanks to A. Belyaev, N. Christensen and A. Pukhov for help with {\tt ClacHEP}, and to G. Huang for comments on the draft.
This work was supported in part by the US DOE Grant DE-FG02-12ER41809, 
the NSF under Award No. EPS-0903806, and 
matching funds from the State of Kansas through the Kansas Technology Enterprise Corporation.

%\begin{verbatim}

%\end{verbatim}

%\bibliographystyle{ws-procs9x6}
%\bibliography{ws-pro-sample}

\end{document}